\documentclass[pdflatex, sn-basic]{sn-jnl}

\usepackage{graphicx}%
\usepackage{multirow}%
\usepackage{amsmath,amssymb,amsfonts}%
\usepackage{amsthm}%
\usepackage{mathrsfs}%
\usepackage[title]{appendix}%
\usepackage{xcolor}%
\usepackage{textcomp}%
\usepackage{manyfoot}%
\usepackage{booktabs}%
\usepackage{algorithm}%
\usepackage{algorithmicx}%
\usepackage{algpseudocode}%
\usepackage{listings}%
\usepackage{float}
\usepackage{arydshln}
\usepackage{lscape}
\usepackage{nicefrac}
\usepackage{lmodern}
\usepackage{bbm}
\usepackage{gensymb}
\usepackage{booktabs}

\usepackage{arydshln}

\theoremstyle{thmstyleone}%
\theoremstyle{thmstyletwo}%

\theoremstyle{thmstylethree}%
\newcommand{\ds}{\displaystyle}

\begin{document}

\title[]{A robust nonparametric test for spatial isotropy in lattice data}


\author*[1]{\fnm{Jana} \sur{Gierse}}\email{gierse@statistik.tu-dortmund.de}

\author[1]{\fnm{Roland} \sur{Fried}}\email{fried@statistik.tu-dortmund.de}

\affil*[1]{\orgdiv{Department of Statistics}, \orgname{TU Dortmund University}, \orgaddress{\street{Vogelpothsweg 87}, \city{Dortmund}, \postcode{44227}, \state{North Rhine-Westphalia}, \country{Germany}}}


\abstract{This paper proposes a robust test for assessing isotropy based on the variogram of spatial data on a two-dimensional regular grid. The test is based on the non-robust subsampling test for isotropy of \cite{Guan2004}, which uses the idea of comparing variogram estimates in different directions at the same distance. The robust test employs robust variogram estimators which are based on estimators of univariate or multivariate scatter and perform well in the presence of isolated or block outliers. Additionally, a different resampling method, called block permutation, is proposed. Compared with the subsampling test, the block permutation test maintains the significance level even for strong dependencies in the data and is robust to outliers. The methods are illustrated by an application to Landsat 8 satellite data, where outlier blocks may occur due to, for example, clouds.}




\newgeometry{bottom = 4cm}
\maketitle
\restoregeometry

\section{Introduction}\label{ch:introduction}

When analysing spatial data, an important step is to investigate the spatial dependencies among the observations. This step is crucial for subsequent analyses, as an incorrectly specified dependency structure can lead to misspecification of the uncertainties and to wrong conclusions in subsequent analyses.
A common measure of spatial dependency is the variogram. For an intrinsically stationary random field ${Z(\boldsymbol{s}), \boldsymbol{s} \in I}$, with index set $I \subset \mathbb{R}^d$, the variogram at a given lag vector $\boldsymbol{h}$ is defined as
	\begin{align*}
		2\gamma(\boldsymbol{h}) = \text{Var}\left(Z(\boldsymbol{s}) - Z(\boldsymbol{s}+\boldsymbol{h})\right),  \mbox{ for }\boldsymbol{s},\boldsymbol{s}+\boldsymbol{h}\in I.
	\end{align*}
Intrinsic stationarity implies that the increments $Z(\boldsymbol{s})-Z(\boldsymbol{s}+\boldsymbol{h})$ are weakly stationary, which means that they have a constant mean and translation-invariant covariance. Often the random field is additionally assumed to be mean stationary, i.e., having a constant mean across the entire domain.

A common simplifying assumption is that the dependency structure is isotropic, which means that the dependence between two measurement points depends only on their distance and not on the direction. This assumption simplifies further analyses, but is questionable in practice.
For example, wind plays an important role in pollen transport in the analysis of crop yields. Consequently, spatial dependence is likely to depend on the dominant direction of the wind \citep{Guan2004}. The assumption of isotropy in such cases can lead to undesirable effects, such as increased kriging variances \citep[p. 87 ff.]{Sherman2011}. A random field that does not exhibit isotropy is referred to as anisotropic. 

In general, a correct specification of the spatial dependency structure improves both the interpretability of the data and the accuracy of predictions \citep{Guan2004}. Various graphical methods have been proposed in the literature to assess anisotropy, such as comparisons of directional variogram estimates or the use of rose diagrams \citep{Matheron1962b, Isaaks1989}. \citet{Diggle1981} and \citet{Banerjee2014} suggest using contour plots of the empirical correlation function, while \citet{Modjeska1983} propose employing empirical estimates of the covariance. However, all of these graphical methods suffer from a lack of objectivity. In general, they are difficult to assess and open to subjective interpretation \citep{Guan2004, Weller2016}.

Therefore, several approaches for formally testing isotropy have been proposed in the literature. \citet{Ecker1999} provide a parametric Bayesian framework for inference on anisotropy parameters. A key drawback of parametric methods is the requirement to specify a model, along with the risk of model misspecification. As an alternative, various nonparametric tests have been developed. An overview of many of these methods is given in \citet{Weller2016}. 
One of these nonparametric tests is proposed by \citet{Guan2004}. The idea is to compare variogram estimates for lag vectors with the same distance but different directions using a contrast-based approach. The test requires an estimate of the covariance matrix of the variogram estimators, which is obtained via subsampling techniques. In addition to an asymptotic version of the test, a resampling-based variant is introduced, particularly suited for small grid sizes; this approach also relies on subsampling. In contrast, \citet{Maity2012} employ kernel-based estimation of the covariance function, while \citet{Petrakis2017} approximate the sampling distribution of the anisotropy parameters.
Whereas these tests are formulated in the spatial domain, the approaches of \citet{Scaccia2002, Scaccia2005, Lu2005, Fuentes2005, Weller2020} are based on spectral methods. Notably, only the test proposed by \citet{Weller2020} is specifically designed for testing isotropy, whereas the others are primarily developed to assess second-order properties of the data, such as symmetry properties of the covariance function.

To the best of our knowledge, none of these tests has been investigated in the presence of outliers. As they are based on non-robust estimators, the resulting tests are presumably not robust either. For example, in Section \ref{ch:simulations} we will see that the test of \citet{Guan2004} fails if outliers occur in the data. To fill this gap, we propose a robust isotropy test, following the lines of the test of \citet{Guan2004}. We replace the non-robust variogram estimator with robust alternatives proposed by \citet{Genton1998a, Gierse2025}. In addition, we suggest a block permutation principle instead of subsampling techniques, both for estimating the covariance of the variogram estimators and for computing p-values.

The remainder of this work is organised as follows. The test proposed by \citet{Guan2004} is reviewed in Section \ref{ch:Guan}. Section \ref{ch:robusttest} describes the proposed robust isotropy test, as well as the block permutation approach. Simulation results for data with and without outliers are presented in Section \ref{ch:simulations}. Section \ref{ch:applications} contains an application to Landsat~8 satellite data. Finally, Section \ref{ch:summary} provides a summary and an outlook.

\section{Isotropy test of \cite{Guan2004}}\label{ch:Guan}

In the following, we assume an intrinsically stationary random field ${Z(\boldsymbol{s}), \boldsymbol{s} \in I}$, defined on a regular grid ($I = \mathbb{Z}^2$). Let $D_n \subset \mathbb{Z}^2$ denote the finite set of grid locations at which observations are available, with $|D_n| = n$. The isotropy test proposed by \cite{Guan2004} compares variogram estimates for lag vectors of equal length but different directions using a contrast-based test. The test employs the classical method-of-moments variogram estimator introduced by \cite{Matheron1962}, which is defined as
	 \begin{align}
	2\widehat{\gamma}_M(\boldsymbol{h}) = \frac{1}{|D(\boldsymbol{h})|} \sum_{\boldsymbol{s}\in D(\boldsymbol{h})} (Z(\boldsymbol{s}) - Z(\boldsymbol{s} + \boldsymbol{h}))^2
    \label{form:math}
	 \end{align}
with $D(\boldsymbol{h}) = \{\boldsymbol{s}: \boldsymbol{s}, \boldsymbol{s} + \boldsymbol{h} \in D_n\}$
being the set of all locations $\boldsymbol{s}$ on the grid for which $\boldsymbol{s}+\boldsymbol{h}$ is also on the grid, and $|D(\boldsymbol{h})|$ being the number of such locations. Under the assumption of mean stationarity, this estimator is unbiased and consistent. \cite{Guan2004} further show that, for a strictly stationary random field, the estimator is asymptotically normally distributed under additional conditions concerning the shape of the field, the strength of spatial dependence, and certain mild moment assumptions.  

Under the assumption of isotropy, the variogram depends only on the length of the lag vector $\boldsymbol{h}$ and can thus be expressed as a function of the Euclidean norm $|\boldsymbol{h}| = \sqrt{\boldsymbol{h}^T \boldsymbol{h}}$. The hypothesis of isotropy can therefore be formulated as $2\gamma(\boldsymbol{h}) = 2\gamma_0(||\boldsymbol{h}||)$
for some isotropic variogram function $2\gamma_0(\cdot)$. The isotropy test proposed by \cite{Guan2004} selects a set $\Lambda$ of lag vectors at which the variogram is estimated and compared. Concentrating on this set, we test the null hypothesis 
\begin{align*}
    H_0: 2\gamma(\boldsymbol{h}_1) = 2\gamma(\boldsymbol{h}_2)~\forall \boldsymbol{h}_1,\boldsymbol{h}_2\in\Lambda \text{ with } ||\boldsymbol{h}_1||=||\boldsymbol{h}_2||.
\end{align*}
Let $\boldsymbol{G} = (2\gamma(\boldsymbol{h}): \boldsymbol{h}\in\Lambda)$ be the vector of the true (unknown) variogram values for all lag vectors of interest, and let $\widehat{\boldsymbol{G}}_{D_n}^\text{M} = (2\widehat{\gamma}_M(\boldsymbol{h}):\boldsymbol{h}\in\Lambda)$ denote the corresponding estimates based on the data observed on the regular grid $D_n$ using the Matheron variogram estimator. Under the null hypothesis, there exists a full-rank matrix $\boldsymbol{A}$ such that $\boldsymbol{A}\boldsymbol{G}\overset{H_0}{=}\boldsymbol{0}$. A contrast test can be used to test $H_0: \boldsymbol{A}\boldsymbol{G}=\boldsymbol{0}$, leading to the test statistic
\begin{align}
    TS_{D_n}^\text{M;sub} = |D_n| \cdot (\boldsymbol{A}\widehat{\boldsymbol{G}}_{D_n}^\text{M})^T(\boldsymbol{A}\widehat{\boldsymbol{\Sigma}}^{\text{sub}}_{\widehat{\boldsymbol{G}}_{D_n}^\text{M}}\boldsymbol{A}^T)^{-1}(\boldsymbol{A}\widehat{\boldsymbol{G}}_{D_n}^\text{M}),
    \label{form:teststatistic}
\end{align}
where $\widehat{\boldsymbol{\Sigma}}_{\widehat{\boldsymbol{G}}_{D_n}^\text{M}}^{\text{sub}}$ denotes an estimate of the covariance matrix of $\widehat{G}_{D_n}^\text{M}$ using a subsampling estimator (see Subsection \ref{ch:subsampling}) and $\boldsymbol{A}^T$ denotes the transpose of $\boldsymbol{A}$. For $\Lambda = \{(1,0)^T, (0,1)^T, (1,1)^T, (-1,1)^T\}$, e.g., the corresponding vector $\boldsymbol{G}$ and a possible contrast matrix $\boldsymbol{A}$ are given by 
\begin{align*}
    \boldsymbol{A} = \begin{pmatrix}
    1 & -1 & 0 & 0 \\ 0 & 0 & 1 -1
    \end{pmatrix},~ \boldsymbol{G} = \begin{pmatrix}
        2\gamma((1,0)^T) \\ 2\gamma((0,1)^T) \\ 2\gamma((1,1)^T) \\ 2\gamma((1,-1)^T)^T
    \end{pmatrix}.
\end{align*}
\cite{Guan2004} show that, for a strictly stationary random field and under the same additional assumptions required for the asymptotic normality of $\widehat{\boldsymbol{G}}_{D_n}^\text{M}$ (e.g., mixing conditions, mild moments assumptions and geometric regularity of $D_n$), the test statistic converges in distribution to a $\chi^2_d$-distribution, $TS_{D_n}^\text{M}\overset{n\rightarrow\infty}{\rightarrow}\chi^2_d$, where $d = \text{rank}(\boldsymbol{A})$ is the number of contrast constraints.

\subsection{Choice of $\Lambda$}
\label{ch:lambda}

An important step in designing the test is to determine $\Lambda$, the set of lags to be tested. The choice of $\Lambda$ depends on the configuration of the data set, the purpose of the study, and the physical or biological phenomenon of interest. For example, if anisotropy in the data arises due to a rotation, it is important to include lag vectors in the southeast–northwest and southwest–northeast directions to detect this anisotropy (see the simulations in Section~\ref{ch:simulations}).

In general, it is advisable to include lag vectors $\boldsymbol{h}$ at short lags, as typically more data are available for them, resulting in more accurate estimates. In addition, the dependence is usually stronger for small lag vectors, making it easier to detect differences. A common choice is $\Lambda = \{(1,0)^T, (0,1)^T, (1,1)^T, (-1, 1)^T\}$. Further details can be found in Section \ref{ch:simulations}, which summarises the simulations.

\subsection{Estimation of the covariance matrix}
\label{ch:subsampling}

The test statistic formulated in  (\ref{form:teststatistic}) requires an estimation of the covariance matrix of the vector $\widehat{\boldsymbol{G}}^\text{M}_{D_n}$ that contains the variogram estimates. \cite{Guan2004} propose a subsampling estimator for this. The sample grid $D_n$ is divided into overlapping blocks that share the same configuration and orientation but are much smaller than $D_n$. To determine these subblocks, a moving subsampling window is applied, and all blocks for which the window is completely contained within $D_n$ are selected. \cite{Shermann1996} shows that a block size of order $l^2(n)$ with $l(n)= c \cdot n^{\frac{1}{2}}$ minimises the mean squared error for estimating the variance of a statistic on a spatial lattice. For each of these subblocks, the variogram is estimated for the predefined set of lags $\Lambda$. The covariance matrix of the variogram estimations is computed using the sample covariance of the estimated variogram vectors from the different subblocks. Let $k_n$ be the number of subblocks,  $D^i_{l(n)}$ the $i-th$ subblock of size $l^2(n)$ and $\widehat{\boldsymbol{G}}_{D^i_{l(n)}}^\text{M}$ the vector of variogram estimations for all lags of $\Lambda$ based on the subblock $D_{l(n)}^i$. The subsampling estimator of the covariance matrix is then given by
\begin{align}
    \widehat{\boldsymbol{\Sigma}}_{\widehat{\boldsymbol{G}}_{D_n}^\text{M}}^{\text{sub}} = \frac{1}{k_n \cdot f_n} \cdot \sum_{i = 1}^{k_n} |D_{l(n)}^{i}|\left( \widehat{\boldsymbol{G}}_{D^i_{l(n)}}^\text{M}-\bar{\boldsymbol{G}}_{n}^\text{M}\right) \left( \widehat{\boldsymbol{G}} _{D^i_{l(n)}}^\text{M}-\bar{\boldsymbol{G}}_{n}^\text{M}\right)^T.
    \label{form:sigma_m}
\end{align}
Here, $f_n = 1-\frac{|D_{l(n)}|}{|D_n|}$ is a finite-sample bias correction factor and $\bar{\boldsymbol{G}}_n^\text{M}$ is the mean of the estimated variogram vectors of the $k_n$ subblocks. Under some additional assumptions (see \citet{Guan2004}), this estimator is $\text{L}_2$-consistent.

In the test statistic, we need the inverse of $\boldsymbol{A}\widehat{\boldsymbol{\Sigma}}_{\widehat{\boldsymbol{G}}_{D_n}^\text{M}}^{\text{sub}}\boldsymbol{A}^T$, see equation \eqref{form:teststatistic}. If $\Lambda$ contains more lag vectors, the problem can occur that the matrix is singular and therefore the inverse does not exist. In such cases, we use some regularisation and add $10\%$ of the average variance to the main diagonal. We thus change the inverse in formula \eqref{form:teststatistic} by the inverse of 
\begin{align}
    \boldsymbol{A}\widehat{\boldsymbol{\Sigma}}_{\widehat{\boldsymbol{G}}_{D_n}^\text{M}}^{\text{sub}}\boldsymbol{A}^T + 0.1 \cdot \frac{\sum_{i=1}^d (\boldsymbol{A}\widehat{\boldsymbol{\Sigma}}_{\widehat{\boldsymbol{G}}_{D_n}^\text{M}}^{\text{sub}}\boldsymbol{A}^T)_{[i,i]}}{d} I_{d\times d}
    \label{form:regularisation}
\end{align}
with $d = \text{rank}(\boldsymbol{A})$, $\boldsymbol{B}_{[i,i]}$ the $[i,i]$-th element of the matrix $\boldsymbol{B}$ and $I_{d\times d}$ the $d\times d$ dimensional identity matrix. 

In general, it is not possible to construct all lag vectors of interest for every location on the grid. Consequently, the estimation of the components in $\boldsymbol{G}$ is based on varying numbers of differences. Furthermore, the number of grid observations $n = |D_n|$ differs from the number of differences used in the estimation of the variogram, $\tilde{n} = |D(\boldsymbol{h})|$. Although the subsampling covariance matrix estimator is consistent, a correction for edge effects is advisable in finite-sample applications \citep{Guan2004}.

For this edge correction, a new spatial process $\{X(\boldsymbol{s}; \boldsymbol{h})\}$ is defined, setting $X(\boldsymbol{s}; \boldsymbol{h}) = [Z(\boldsymbol{s}) - Z(\boldsymbol{s} + \boldsymbol{h})]^2$. In addition, the set $D_n(\Lambda)$ is defined to contain only the grid points for which every lag in the set $\Lambda$ can be constructed,
$D_n(\Lambda) = \bigcap_{\boldsymbol{h}\in\Lambda}D_n(\boldsymbol{h})$. The variogram is then estimated using the set $D_n(\Lambda)$ and the newly defined spatial process \citep{Guan2004}
\begin{align*}
    2\tilde{\gamma}^\text{M}_{D_n}(\boldsymbol{h}) = \frac{1}{|D_n(\Lambda)|}\sum_{s\in D_n(\Lambda)}X(\boldsymbol{s}, \boldsymbol{h}).
\end{align*}
For large grids, this estimator is approximately equal to the one based on the set $D_n(\boldsymbol{h})$, see \eqref{form:math}. For the edge effect corrected version of the subsampling covariance estimator,  $\widehat{\boldsymbol{G}}^\text{M}_{D^i_{l(n)}}$ is replaced by $\tilde{\boldsymbol{G}}^\text{M}_{D^i_{l(n)}} = \{2\tilde{\gamma}^\text{M}_{D^i_{l(n)}}(\boldsymbol{h}): \boldsymbol{h}\in\Lambda\}$ with $2\tilde{\gamma}^\text{M}_{l(n);i}(\boldsymbol{h})$ being the sample mean of the new process in $D_{l(n)}^i\subset D_n(\Lambda)$.

\subsection{Subsampling Test}

As mentioned above, the test statistic converges to a $\chi^2$-distribution under $H_0$. However, according to \citet{Guan2004} this convergence is quite slow. Therefore, they propose a subsampling test, suitable in particular for small grids. The test statistic is calculated for the same subblocks as used in the estimation of the covariance matrix. Under $H_0$, the sampling distribution of the test statistic $TS_{D_n}^\text{M}$ can be approximated by
\begin{align*}
    \widehat{F}_{n,l(n)}(x) = \frac{1}{k_n}\sum_{i=1}^{k_n}\mathbbm{1}(TS_{D^i_{l(n)}}^\text{M;sub}\leq x)
\end{align*}
with $TS_{D^i_{l(n)}}^\text{M;sub}$ denoting the test statistic for the $i-th$ subblock $D_{l(n)}^{i}$ and $\mathbbm{1}(\cdot)$ the indicator function. The resulting p-value is 
\begin{align*}
    p_{\text{sub}} = \frac{1}{k_n}\sum_{i=1}^{k_n}\mathbbm{1}(TS_{D^i_{l(n)}}^\text{M;sub} \geq TS^\text{M;sub}_{D_n}).
\end{align*}

\section{Robust test for isotropy}\label{ch:robusttest}

The test described above is based on the Matheron variogram estimator, which is known to be non-robust and highly sensitive to outliers \citep{Genton1998a}. Additionally, the proposed subsampling approach uses variogram estimation on many small subsamples. These two issues make the test non-robust (see Section \ref{ch:sim_out}). Therefore,  we propose robustifications of the test statistic in Subsection \ref{ch:robtest}, and in Subsection \ref{ch:blockpermutation} we present an alternative to the subsampling approach.

\subsection{Robust test statistics}
\label{ch:robtest}
In order to improve the robustness of the test, we replace Matheron's classical variogram estimator with a robust alternative. \citet{Genton1998a} proposed a robust variogram estimator based on the $Q_n$ scale estimator introduced by \citet{Rousseeuw1993}, which combines good robustness properties with considerable Gaussian efficiency. This variogram estimator has been shown to perform well in the presence of isolated outliers \citep{Lark2000, Kerry2007a}. It is defined as 
\begin{align*}
	2\widehat{\gamma}_\text{G}(\boldsymbol{h}) = \left(c \cdot \left(|V_i(\boldsymbol{h}) - V_j(\boldsymbol{h})|: i<j\right)_{(k)}\right)^2
	\end{align*} 
where $k = \begin{pmatrix}\left[\nicefrac{|D(\boldsymbol{h})|}{2}\right] + 1 \\ 2\end{pmatrix}$,  $[a]$ denotes the integer part of $a$, $(\cdot)_{(k)}$ denotes the $k$-th order statistic and
$c$ is a consistency factor, which  approximately equals $2.22$ for large Gaussian samples. Furthermore, $V_i(\boldsymbol{h})$ is the $i$-th element of the set $V(\boldsymbol{h}) = \{Z(\boldsymbol{s}) - Z(\boldsymbol{s} + \boldsymbol{h}): \boldsymbol{s}\in D(\boldsymbol{h})\}$ containing all differences for locations with distance $\boldsymbol{h}$. This estimator exhibits robustness, with a spatial breakdown point of at least $25\%$ in the case of two-dimensional data. The spatial breakdown point is defined with respect to the most adverse spatial arrangement of the outliers and accounts for the fact that in spatial settings the location of outliers plays a crucial role \citep{Genton1998, Lark2008}.
A robust isotropy test based on the Genton variogram estimator can then be defined by replacing $\widehat{\boldsymbol{G}}^\text{M}_{D_n}$ with $\widehat{\boldsymbol{G}}^\text{G}_{D_n} = \{2\widehat{\gamma}_\text{G}(\boldsymbol{h}): \boldsymbol{h}\in\Lambda\}$ in the test statistic \eqref{form:teststatistic} and $\widehat{\boldsymbol{G}}^\text{M}_{D^i_{l(n)}}$ with  $\widehat{\boldsymbol{G}}^\text{G}_{D^i_{l(n)}}$ in the covariance estimator \eqref{form:sigma_m}.

If the outliers occur spatially aggregated as a block rather than isolated, the Genton variogram estimator is strongly affected by such outlier blocks and is therefore no longer the best choice. 
Using the highly robust Minimum Covariance Determinant (MCD) estimator of multivariate location and scatter introduced by \citet{Rousseeuw1985},
\citet{Gierse2025} propose a directional variogram estimator, called MCD.diff, which is specifically designed to handle outlier blocks. Their estimator jointly estimates the directional variogram in one of the four principal directions (S–N, E–W, SW–NE, SE–NW) for different lags. 
This can be easily expanded for estimation in an arbitrary direction by constructing vectors of the form given in equation  \eqref{form:W}, where $\boldsymbol{h}_1, \ldots, \boldsymbol{h}_{h_{\max}}$ are vectors in the desired direction.

The MCD selects a subset of size $k(n)$ from the total sample of size $n$ such that the determinant of the estimated covariance matrix is minimal, where $\lfloor \nicefrac{n + p + 1}{2} \rfloor \leq k(n) \leq n$ with $p$ the dimension of the vectors. The location is estimated as the mean of these $k(n)$ data points, and the scatter is estimated as their empirical covariance matrix, multiplied by a consistency factor. To improve efficiency, a reweighting step is often included. Various algorithms exist for computing this estimator, such as the FASTMCD algorithm proposed by \citet{Rousseeuw1999}. For more details on the MCD estimator, see \citet{Hubert2010} or \citet{Hubert2018}. 

The MCD-based directional variogram estimator MCD.diff constructs vectors of differences between the observations so that the variogram is represented on the main diagonal of the variance–covariance matrix of these vectors. The reweighted MCD is then used to estimate this covariance matrix.     
If $\boldsymbol{h}_1, \ldots , \boldsymbol{h}_{h_{\max}}$ are lag vectors in the same direction for which we wish to estimate the variogram jointly, vectors of the following form are constructed:
\begin{align}
		\boldsymbol{W} = \begin{pmatrix}
			Z(\boldsymbol{s}) - Z(\boldsymbol{s}+\boldsymbol{h}_1) \\
			Z(\boldsymbol{s}) - Z(\boldsymbol{s}+\boldsymbol{h}_2) \\
			\vdots \\
			Z(\boldsymbol{s}) - Z(\boldsymbol{s}+\boldsymbol{h}_{h_{\max}})
		\end{pmatrix}.
        \label{form:W}
\end{align}
For estimation in the S–N direction, the lag vectors take the form $\boldsymbol{h}_l = (0,l)^{T}$, $l = 1, \ldots, h_{\max}$, and for estimation in the SW–NE direction, we choose $\boldsymbol{h}_l = (l,l)^{T}$, $l = 1, \ldots, h_{\max}$.
The estimated variogram vector $(2\widehat{\gamma}_{MCD}(\boldsymbol{h}_1), \ldots , 2\widehat{\gamma}_{MCD}(\boldsymbol{h}_{h_{\max}}))^T$ corresponds to the diagonal of the estimated covariance matrix of all possible vectors $\boldsymbol{W}$, obtained using the reweighted MCD covariance estimator. 
A robust isotropy test can be constructed by replacing the Matheron variogram estimator in the test statistic \eqref{form:teststatistic} and in the subsampling covariance estimator \eqref{form:sigma_m} with the MCD.diff variogram estimator. When using the subsampling covariance estimator in combination with MCD.diff, we must ensure that the subsamples are not too small and that enough vectors can be built for estimation. At least $2h_{\max}+1$ vectors are needed for the joint estimation of $h_{\max}$ different lag vectors in one direction. 

\cite{Gierse2025} show that, for small grid sizes, the MCD.diff variogram estimator requires an additional correction factor depending on the dimension of the vectors $h_{\max}$. In the isotropy test used here, in general only one lag is estimated in each direction (see Sections \ref{ch:lambda} and \ref{ch:simulations}). Simulations not shown here indicate that, in this case, the additional correction factors are required, even for the small subsamples in the subsampling approach.

The subsampling estimator of the covariance of the estimated variogram vector partitions the data into small blocks and estimates the variogram for the lags of interest within each block. 
This approach is not robust if the classical empirical covariance estimator is applied, since a few  individual variogram estimates which are distorted by outliers are sufficient to cause a substantial bias for the covariance estimate.
Because the blocks are small, already a few outliers cause a large proportion of outliers in an individual block and thus a strong bias even for robust variogram estimators.
Therefore, we use the robust MCD  for the subsampling covariance estimation.

\subsection{Block permutation}\label{ch:blockpermutation}

\citet{Guan2004} propose subsampling for the estimation of the covariance matrix of the variogram estimator and for the determination of the p-value. However, the subsampling approach leads to liberal tests if the spatial dependencies in the data are strong (see Subsection \ref{ch:sim_oA}). Furthermore, as mentioned above, already a few outliers can strongly influence even robust variogram estimators when applied to rather small subsamples. This is particularly true in the case of an outlier block. Therefore, the subsampling approach is not robust (see Subsection \ref{ch:sim_out}). 

We instead suggest a block permutation approach, which can be used for covariance estimation and for determination of the p-value.  
In contrast to the subsampling approach, we divide the grid into non-overlapping blocks, which should have the same configuration and orientation as the original grid. Under the null hypothesis of isotropy, the dependency structure of the data does not depend on the direction. Therefore, the dependency structure of the data in each block remains unchanged after a 90-degree rotation. The idea is thus to randomly decide for each block whether it is rotated by 90 degrees or left unchanged. This results in a new dataset in which some blocks remain unchanged, while others are rotated. Problems arise due to the differences between data points from rotated and unrotated blocks. To avoid this, we only use differences within the same block to obtain a variogram estimate for every lag vector in $\Lambda$. This procedure is repeated $B$ times, leading to $B$ permutated data sets $D_n^{1}, \ldots , D_n^{B}$. 

\begin{figure}[ht]
\begin{minipage}{0.48\textwidth}
    \centering
    \includegraphics[width=1\linewidth]{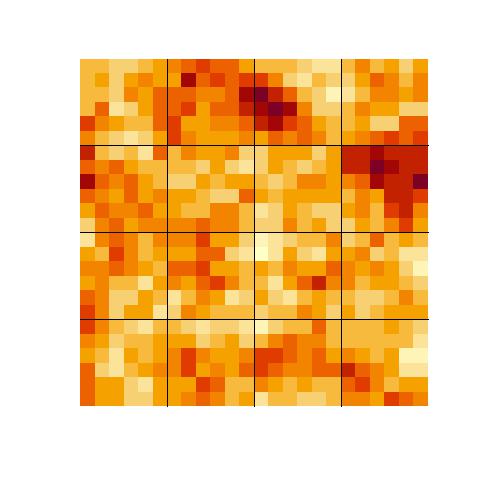}
\end{minipage}
\begin{minipage}{0.48\textwidth}
     \centering
    \includegraphics[width=1\linewidth]{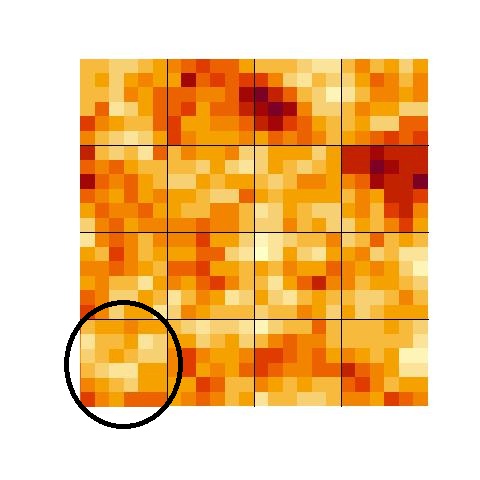}
\end{minipage}
    \caption{Example of a block permutated dataset (right side), where compared to the original dataset (left side) only the block in the left corner is rotated by 90 degrees.}
    \label{fig:blockpermutation}
\end{figure}

Figure \ref{fig:blockpermutation} shows, on the right side, an exemplary data set after block permutation. Compared to the original data set on the left side, here only the block in the lower left corner is rotated by 90 degrees, while all other blocks remain unchanged.

From the resulting $B$ estimates, the covariance of the variogram estimation can be calculated using a covariance estimator such as the robust MCD. If singularity problems occur, regularisation is used as in the subsampling approach (see equation \eqref{form:regularisation}). For the test decision, the test statistic is computed for each of the $B$ data sets, and the p-value is then determined based on these values, e.g.,
\begin{align*}
    p_{\text{block}} = \frac{1}{B}\sum_{i=1}^{B}\mathbbm{1}(TS_{D^{i}_n}^\text{block} \geq TS^\text{block}_{D_n}).
\end{align*}
where $TS_{D_{n}^i}^\text{block}$ is the test statistic for the $i-$th block permutated data set $D_{n}^i$ and $TS_{D_n}^{\text{block}}$ the test statistic based on the original data. 

In preliminary simulations, the differences with respect to the use of different block sizes were small, with smaller blocks leading to somewhat conservative tests. Therefore, we suggest using a moderate block size, e.g., blocks of size $8 \times 8$ for a $40 \times 40$ grid, which results in $25$ non-overlapping blocks and $2^{25}$ different possible permutated data sets. See also Table \ref{tab:oA_40} in the Appendix.

\section{Simulation results}\label{ch:simulations}

In the following, we evaluate the subsampling test and the block permutation test using three different variogram estimators (Matheron, Genton, and MCD.diff) via simulations. In these simulations, we consider a mean-stationary Gaussian random field  $Z(\boldsymbol{s})$ on a two-dimensional $24\times 24$ dimensional regular grid. To investigate the behaviour of the tests under the alternative of anisotropy, we simulate random fields with geometric anisotropic variograms, which can be transformed into isotropic behaviour by clockwise rotation with an angle $\theta$ and rescaling of the coordinate axes; see Figure \ref{fig:aniso} in the appendix for an illustration. 

 The variogram, the 
rotation matrix $\boldsymbol{R}$ and the rescaling matrix $\boldsymbol{T}$ are given by 
\begin{align*}2\gamma(\boldsymbol{h}) = 2\gamma_{0}\left(\sqrt{\boldsymbol{h}^{T}\boldsymbol{R}^{T}\boldsymbol{T}^{T}\boldsymbol{T}\boldsymbol{R}\boldsymbol{h}}\right),~~
    \boldsymbol{R} = \begin{pmatrix}
        \cos(\theta) & -\sin(\theta) \\
        \sin(\theta) & \cos(\theta)
    \end{pmatrix}
    \boldsymbol{T} = \begin{pmatrix}
        1 & 0 \\
        0 & \frac{1}{b}
    \end{pmatrix}, 
    \end{align*}
with an isotropic variogram $2\gamma_0(\cdot)$ \citep[pp. 91]{Sherman2011}. 
In the simulations, we use a spherical variogram, i.e., 
	\begin{align*}
	    \gamma_{0}(\boldsymbol{h}) = 
	    \beta \left( \frac{3||\boldsymbol{h}||}{2r}-\frac{||\boldsymbol{h}||^3}{2r^3}\right)   	    \mathbbm{1}(0 < ||\boldsymbol{h}|| < r)+
	    \beta \mathbbm{1}(||\boldsymbol{h}|| \geq r).
	\end{align*}
with a sill $\beta = 1$. The range $r$ of a variogram is the distance at which the variogram reaches its sill, indicating that observations are no longer spatially correlated beyond this point. Therefore, the range is a measure of the strength of spatial dependence. We investigate random fields with a short ($r = 2$), moderate ($r = 5$), and a large range ($r = 8$). A visualisation of the true isotropic variogram with range $5$ can be found in the appendix (Figure \ref{fig:aniso}). 

Table \ref{tab:anisotropy} contains the parameter combinations used in the simulations to generate isotropic or anisotropic random fields. The first combination ($\theta = 0, b = 1$) corresponds to an isotropic random field, i.e., a model under the null hypothesis. The second and third combination lead to geometric anisotropic random fields with (only) scaling by different ratios, while the other two combinations lead to geometric anisotropic random fields with different ratios and an additional rotation. 

\begin{table}[ht]
    \begin{tabular}{c|ccccc}
        $\theta$ & 0 & 0 & 0  & $\nicefrac{\ds \pi}{\ds 4}$ & $\nicefrac{\ds \pi}{\ds 4}$ \\ 
     $b$ & 1 & $\sqrt{2}$ & $ 2$ & $\sqrt{2}$ & 2\\
    \end{tabular}
    \caption{Parameters of the geometric anisotropy ($\theta$: rotation angle, $b$: scaling factor) used in the simulations.}
    \label{tab:anisotropy}
\end{table}

For the subsampling approach, we use, as suggested in \cite{Guan2004}, subsamples of the size $l^2(n)$ with $ l(n)= n^{\frac{1}{2}}$. For the $24\times 24$ grid, this results in overlapping subsamples of size $5\times 5$. Furthermore, we apply the suggested edge correction. In the block permutation approach, a block size of $6\times 6$ is used, resulting in 16 nonoverlapping blocks, and $B=1000$ permutated samples are generated. In both the subsampling and the block permutation approach, we use the robust MCD estimator instead of the ordinary sample covariance to estimate the covariance matrix of the variogram estimator. In simulations not shown here, we observed that the differences are only small with the robust version yielding a slightly better performance in the presence of outliers.

We investigate three choices of the set of lags $\Lambda$ in the following. The first set only compares one-step estimations in the north-south and in the east-west direction. The other two lag sets additionally include comparisons in other directions,
\begin{align*}
    \Lambda_1 = \left\{\begin{pmatrix}
        1 \\ 0
    \end{pmatrix}, \begin{pmatrix}
        0 \\ 1
    \end{pmatrix}\right\}, ~
        \Lambda_2 = \left\{\begin{pmatrix}
        1 \\ 0
    \end{pmatrix}, \begin{pmatrix}
        0 \\ 1
    \end{pmatrix}, \begin{pmatrix}
        1 \\ 1
    \end{pmatrix},\begin{pmatrix}
        1 \\ -1
    \end{pmatrix}\right\}, \\
        \Lambda_3 = \left\{\begin{pmatrix}
        1 \\ 0
    \end{pmatrix}, \begin{pmatrix}
        0 \\ 1
    \end{pmatrix}, \begin{pmatrix}
        1 \\ 1
    \end{pmatrix},\begin{pmatrix}
        1 \\ -1
    \end{pmatrix},
    \begin{pmatrix}
        2 \\ 1
    \end{pmatrix},
    \begin{pmatrix}
        -1 \\ 2
    \end{pmatrix},
    \begin{pmatrix}
        1 \\ 2
    \end{pmatrix},
    \begin{pmatrix}
        -2 \\ 1
    \end{pmatrix}\right\}.
\end{align*}
The corresponding contrast matrices are given by
\begin{align*}
    \boldsymbol{A}_1 = \begin{pmatrix}
        1 & -1 
    \end{pmatrix}, \boldsymbol{A}_2 = \begin{pmatrix}
        1 & -1 & 0 & 0 \\
        0 & 0 & 1 & -1 
    \end{pmatrix}, \boldsymbol{A}_3 = \begin{pmatrix}
            1 & -1 & 0 & 0 & 0 & 0 & 0 & 0 & \\
        0 & 0 & 1 & -1 & 0 & 0 & 0 & 0 \\
        0 & 0 & 0 & 0 & 1 & -1 & 0 & 0 \\
        0 & 0 & 0 & 0 & 0 & 0 &  1 & -1
    \end{pmatrix}.
\end{align*}
The simulations are performed on a HPC cluster using the software \texttt{R} 4.4.2 \citep{r2024}. The R-package \texttt{geoR} \citep{geoR2025} is used for the simulation of Gaussian random fields. For all settings 1000 samples are simulated and a significance level of $5\%$ is used. All tests are performed using the R-package \texttt{RobVario} available on github (\url{https://github.com/JGierse/RobVario.git}).  

Table \ref{tab:runtimes} reports computing times (in seconds) on an HPC cluster equipped with two Intel Xeon E5-2640v4 processors, for regular grids of size $24\times 24$ and $40\times 40$. The MCD.diff estimator requires substantially more computation time than the other two estimators, which exhibit similar runtimes. For the subsampling approach, a slight influence of the lag set $\Lambda$ can be observed for the Matheron and Genton estimators, and this effect is more pronounced for the MCD.diff estimator. In general, and as expected, a larger number of lags in $\Lambda$ and a larger grid size increase both computation time. The reported computation times for the block permutation approach are based on 1000 permutation samples. These runtimes are considerably higher than those of the subsampling approach. Moreover, the influence of the lag set $\Lambda$ is more pronounced in this case for the smaller grid. For the larger grid no clear relationship between runtime and lag set can be seen.

\begin{table}[ht]
    \centering
    \begin{tabular}{cc|ccc|ccc}
         & & \multicolumn{3}{c}{Subsampling} & \multicolumn{3}{c}{Block permutation} \\
         size & & Matheron & Genton & MCD.diff & Matheron & Genton & MCD.diff \\ \hline
         \multirow{3}{*}{$24\times 24$} & $\Lambda_1$ & 1.39 & 1.43 & 2.42 & 20.86 & 21.84 & 24.22  \\
                                        & $ \Lambda_2$ & 1.43 & 1.56 & 3.10 & 35.03 & 38.28 & 42.94 \\
                                        & $\Lambda_3$ & 1.66 & 1.73 & 4.00 & 59.82 & 69.49 & 73.06 \\ \hline
        \multirow{3}{*}{$40\times 40$} & $\Lambda_1$ & 27.58 & 27.41 & 30.99 & 91.70 & 143.88 & 265.44 \\
                                       & $\Lambda_2$ & 28.47 & 27.34 & 47.56 & 99.90 & 134.25 & 256.60 \\
                                       & $\Lambda_3$ & 31.11 & 28.60 & 48.90 & 91.86 & 137.97 & 224.88
    \end{tabular}
    \caption{Computing times in seconds on an HPC cluster with two Intel Xeon E5-2640v4 processors. In the block permutation approach 1000 permutations are used.}
    \label{tab:runtimes}
\end{table}

\subsection{Gaussian data}
\label{ch:sim_oA}

First, we consider simulation results for Gaussian random fields without outliers, as described at the beginning of Section \ref{ch:simulations}. Table \ref{tab:oA} shows the observed percentages of rejections of the null hypothesis for different ranges $r$ of the variogram and the three different lag sets $\Lambda_1, \Lambda_2, \Lambda_3$ in case of the subsampling approach (upper part of the table) and the block permutation approach (lower part). The columns represent different parameter settings of the rotation angle $\theta$ and the ratio $b$ of the geometric anisotropy. The first parameter combination $\theta = 0$ and $b= 1$ corresponds to an isotropic random field and thus represents simulation results for the type one error.

\begin{table}[ht]
    \centering
    \footnotesize
    \begin{tabular}{cc|rrr:rrr:rrr:rrr:rrr}
     & &  \multicolumn{15}{c}{Subsampling} \\
      &  & \multicolumn{3}{c}{$\theta = 0; b = 1$} & \multicolumn{3}{c}{$\theta = 0; b = \sqrt{2}$} & \multicolumn{3}{c}{$\theta = 0; b = 2$} & \multicolumn{3}{c}{$\theta = \nicefrac{\pi}{4}; b = \sqrt{2}$} & \multicolumn{3}{c}{$\theta = \nicefrac{\pi}{4}; b = 2$} \\
     $r$ & estimator & $\Lambda_1$ & $\Lambda_2$ & $\Lambda_3$ & $\Lambda_1$ & $\Lambda_2$ & $\Lambda_3$ & $\Lambda_1$ & $\Lambda_2$ & $\Lambda_3$ & $\Lambda_1$ & $\Lambda_2$ & $\Lambda_3$ & $\Lambda_1$ & $\Lambda_2$ & $\Lambda_3$  \\ \hline  
     \multirow{3}{*}{2} & Matheron & 5 & 5 & 7 & 92 & 85 & 73 & 100 & 100 & 99  & 6 & 68 & 57 & 4 & 100 & 97 \\
                        & Genton & 2 & 1 & 2 & 71 & 54 & 38 & 100 & 99 & 91 & 2 & 36 & 24 & 1 & 95 & 80 \\
                        & MCD.diff & 10 & 12 & 16  & 75 & 67 & 61 & 99 & 98 & 94 & 12 & 51 & 47  & 8 & 96 & 88 \\ \hdashline
    \multirow{3}{*}{5} & Matheron &  6 & 6 & 7  & 91 & 80 & 65 & 100 & 100 & 99 & 6 & 70 & 53 & 6 & 100 & 96 \\
                       & Genton & 3 & 5 & 7 & 79 & 66 & 56 & 100 & 99 & 97  & 5 & 63 & 57 & 5 & 99 & 96 \\
                       & MCD.diff &  13 & 20 & 29 & 80 & 75 & 72 & 100 & 99 & 98  & 16 & 73 & 72  & 16 & 98 & 98 \\ \hdashline
    \multirow{3}{*}{8} & Matheron &  7 & 8 & 11 & 92 & 81 & 66 & 100 & 100 & 98  & 8 & 70 & 53 & 6 & 99 & 95 \\
                       & Genton & 7 & 12 & 24 & 85 & 77 & 71 & 100 & 100 & 98 & 9 & 74 & 69 & 6 & 99 & 97   \\
                       & MCD.diff & 17 & 27 & 47 & 85 & 82 & 83 & 100 & 99 & 98 & 18 & 81 & 82  & 16 & 99 & 99 \\
    \end{tabular}
    
    \vspace{0.2cm}
      \begin{tabular}{cc|rrr:rrr:rrr:rrr:rrr}
     & &  \multicolumn{15}{c}{Block permutation} \\
      &  & \multicolumn{3}{c}{$\theta = 0; b = 1$} & \multicolumn{3}{c}{$\theta = 0; b = \sqrt{2}$} & \multicolumn{3}{c}{$\theta = 0; b = 2$} & \multicolumn{3}{c}{$\theta = \nicefrac{\pi}{4}; b = \sqrt{2}$} & \multicolumn{3}{c}{$\theta = \nicefrac{\pi}{4}; b = 2$} \\
     $r$ & estimator & $\Lambda_1$ & $\Lambda_2$ & $\Lambda_3$ & $\Lambda_1$ & $\Lambda_2$ & $\Lambda_3$ & $\Lambda_1$ & $\Lambda_2$ & $\Lambda_3$ & $\Lambda_1$ & $\Lambda_2$ & $\Lambda_3$ & $\Lambda_1$ & $\Lambda_2$ & $\Lambda_3$  \\   \hline
     \multirow{3}{*}{2} & Matheron & 3 & 3 & 5 & 91 & 79 & 65 & 100 & 100 & 98 & 5 & 52 & 56 & 4 & 98 & 95 \\
                        & Genton & 3 & 4 & 4 & 84 & 71 & 53  & 100 & 100 & 98 & 4 & 44 & 39 & 4 & 98 & 94 \\
                        & MCD.diff &  3 & 4 & 1 & 65 & 49 & 19 & 99 & 98 & 71  & 5 & 28 & 11  & 3 & 90 & 42 \\ \hdashline
    \multirow{3}{*}{5} & Matheron & 5 & 5 & 4 & 90 & 80 & 66 & 100 & 100 & 99 & 6 & 54 & 50 & 5 & 98 & 96 \\
                       & Genton & 4 & 5 & 4 & 84 & 73 & 55 & 100 & 100 & 99 & 6 & 50 & 37 & 5 & 98 & 94 \\
                       & MCD.diff & 5 & 5 & 1 & 67 & 54 & 16 & 100 & 98 & 56  & 4 & 37 & 7 & 4 & 92 & 24 \\ \hdashline
    \multirow{3}{*}{8} & Matheron & 6 & 5 & 6 & 90 & 79 & 68 & 100 & 100 & 99 & 6 & 54 & 52 & 5 & 98 & 96 \\
                       & Genton & 6 & 5 & 4  & 85 & 74 & 56 & 100 & 100 & 99 & 5 & 50 & 36 & 4 & 99 & 93 \\
                       & MCD.diff &  4 & 5 & 1 & 68 & 53 & 14 & 100 & 98 & 54  & 5 & 38 & 6  & 4 & 94 & 26 
    \end{tabular}
    \caption{Simulated percentages of rejections for Gaussian data without outliers and variograms with different ranges $r$ in case of of a $24\times 24$ grid. The upper part of the table contains the results for the subsampling approach and the lower part for the block permutation approach.}
    \label{tab:oA}
\end{table}

Under the null hypothesis, the empirical rejection rates of the block permutation test are close to or below the nominal significance level in all cases. In contrast, the subsampling approach exceeds the nominal significance level when strong spatial dependence is present in the data. This drawback becomes more pronounced as the lag set includes more vectors (i.e., $\Lambda_2$, $\Lambda_3$) and when using robust estimators (Genton or MCD.diff). Subsampling in combination with the MCD.diff estimator fails to maintain the significance level even when the dependence is weak, and this issue already occurs for $\Lambda_1$. The block permutation test based on the MCD.diff estimator with the lag set $\Lambda_3$ also has problems: in contrast to the subsampling test, the type one error rate of the block permutation test with this configuration is very small, which implies a substantial loss of power. 

We observe, as expected, that a higher ratio $b$ generally leads to larger power for all estimators and both approaches (except for  block permutation combined with the MCD.diff estimator and $\Lambda_3$). Moreover, the empirical rejection rates are often close to $100\%$.  Therefore, in the following, we focus on the differences between the tests in case of the smaller ratio $b = \sqrt{2}$.

Table \ref{tab:oA} highlights the importance of choosing the lag set $\Lambda$ properly. If anisotropy arises not only from scaling but also from rotation, it is important to include lag vectors not only in the east–west and north–south directions. This is particularly evident here for $\theta = \nicefrac{\pi}{4}$ and $b = 2$: whereas for $\Lambda_2$ and $\Lambda_3$, which include more directions, the power is nearly one, the power for $\Lambda_1$ is nearly 0. However, if the anisotropy is only due to different scalings in the main directions, including vectors for the other directions can result in a loss of power (see the results for $\Lambda_2$ and $\Lambda_3$ in case of $\theta=0$ and $b=\sqrt{2}$).

We observe that the Matheron estimator leads to the highest power. This is expected since this variogram estimator performs best for Gaussian data without outliers \citep{Gierse2025}. Furthermore, we observe no substantial differences between the two testing approaches when using this estimator, nor between the power values across different ranges. 
Comparing the two robust estimators, we find that the Genton estimator delivers higher power in nearly all cases. In the subsampling approach, the power of the robust estimators depends on the variogram range. This is different for the block permutation approach, where the range has no noticeable influence on the power. If the dependence in the data is weak,  the block permutation approach yields higher power for the robust estimators than the subsampling approach. If the spatial dependence is moderate, the two approaches perform similarly well. In contrast, if the dependence is strong, the subsampling approach leads to higher power for the robust estimators. However, it should be kept in mind that the subsampling approach leads not only to an increase in power for larger ranges, but also to an increase of the type I error rate. Table \ref{tab:oA_size} shows the size-corrected power of the tests. The differences between the size-corrected powers of the two approaches are often small, particularly if the anisotropy ratio  $b=2$ is large. 
For the smaller anisotropy ratio, 
the size-corrected power observed is higher for block permutation than for subsampling if the range of the dependences is large and  the anisotropy angle is $0$. In contrast, the subsampling approach yields higher size-corrected power  when anisotropy arises not only from scaling but also from rotation.
However, we need to keep in mind that subsampling has difficulties with keeping the significance level, particularly if the range of dependence is large. 

\begin{table}[ht]
    \centering
    \footnotesize
    \begin{tabular}{cc|rrr:rrr:rrr:rrr}
     & &  \multicolumn{12}{c}{Subsampling} \\
      &  & \multicolumn{3}{c}{$\theta = 0; b = \sqrt{2}$} & \multicolumn{3}{c}{$\theta = 0; b = 2$} & \multicolumn{3}{c}{$\theta = \nicefrac{\pi}{4}; b = \sqrt{2}$} & \multicolumn{3}{c}{$\theta = \nicefrac{\pi}{4}; b = 2$} \\
     $r$ & estimator  & $\Lambda_1$ & $\Lambda_2$ & $\Lambda_3$ & $\Lambda_1$ & $\Lambda_2$ & $\Lambda_3$ & $\Lambda_1$ & $\Lambda_2$ & $\Lambda_3$ & $\Lambda_1$ & $\Lambda_2$ & $\Lambda_3$  \\ \hline 
     \multirow{3}{*}{2} & Matheron & 94 & 90 & 74 & 100 & 100 & 100 & 5 & 73 & 56  & 5 & 100 & 99 \\
                        & Genton & 91 & 82 & 66 & 100 & 100 & 100 & 6 & 66 & 47  & 6 & 100 & 98 \\
                        & MCD.diff &  71 & 57 & 46 & 100 & 99 & 95 & 5 & 38 & 34 & 4 & 96 & 89 \\ \hdashline
    \multirow{3}{*}{5} & Matheron & 91 & 80 & 56 & 100 & 100 & 100 & 6 & 66 & 42 & 6 & 100 & 95  \\
                       & Genton & 88 & 75 & 55 & 100 & 100 & 99 & 7 & 69 & 52 & 7 & 100 & 97 \\
                       & MCD.diff & 72 & 53 & 39 & 100 & 98 & 93  & 7 & 53 & 40  & 8 & 98 & 91  \\ \hdashline
    \multirow{3}{*}{8} & Matheron & 89 & 77 & 51 & 100 & 100 & 98 & 5 & 63 & 42 & 4 & 100 & 94  \\
                       & Genton & 87 & 66 & 37 & 100 & 100 & 93 & 6 & 61 & 37 & 4 & 99 & 90 \\
                       & MCD.diff & 71 & 51 & 30 & 100 & 97 & 81  & 6 & 49 & 32 & 4 & 96 & 82 \\
    \end{tabular}
    
    \vspace{0.2cm}
      \begin{tabular}{cc|rrr:rrr:rrr:rrr}
     & &  \multicolumn{12}{c}{Block permutation} \\
      &  &  \multicolumn{3}{c}{$\theta = 0; b = \sqrt{2}$} & \multicolumn{3}{c}{$\theta = 0; b = 2$} & \multicolumn{3}{c}{$\theta = \nicefrac{\pi}{4}; b = \sqrt{2}$} & \multicolumn{3}{c}{$\theta = \nicefrac{\pi}{4}; b = 2$} \\
     $r$ & estimator & $\Lambda_1$ & $\Lambda_2$ & $\Lambda_3$ & $\Lambda_1$ & $\Lambda_2$ & $\Lambda_3$ & $\Lambda_1$ & $\Lambda_2$ & $\Lambda_3$ & $\Lambda_1$ & $\Lambda_2$ & $\Lambda_3$  \\   \hline
     \multirow{3}{*}{2} & Matheron & 94 & 86 & 69 &  100 & 100 & 99 & 6 & 64 & 60 & 6 & 99 & 96 \\
                        & Genton & 88 & 81 & 59 & 100 & 100 & 99 & 6 & 56 & 43 & 5 & 99 & 96 \\
                        & MCD.diff & 71 & 55 & 44 & 100 & 99 & 96  & 7 & 35 & 31  & 4 & 93 & 85 \\ \hdashline
    \multirow{3}{*}{5} & Matheron & 90 & 81 & 72 & 100 & 100 & 99  & 5 & 54 & 54  & 4 & 97 & 96 \\
                       & Genton &  87 & 75 & 62 & 100 & 100 & 100 & 7 & 52 & 43  & 5 & 98 & 96 \\
                       & MCD.diff &  69 & 54 & 43  & 100 & 98 & 95 & 5 & 38 & 22  & 5 & 93 & 74 \\ \hdashline
    \multirow{3}{*}{8} & Matheron & 88 & 76 & 61 & 100 & 100 & 98 & 5 & 52 & 45 & 5 & 97 & 94 \\
                       & Genton & 84 & 76 & 60  & 100 & 100 & 100 & 5 & 52 & 42 & 4 & 99 & 95 \\
                       & MCD.diff &   72 & 51 & 46 & 100 & 98 & 95 & 6 & 39 & 24  & 4 & 94 & 80
    \end{tabular}
    \caption{Empirical size-corrected power (in percent)  for Gaussian data in case of variograms with different ranges $r$ and a $24\times 24$ grid. The upper part contains the results for the subsampling approach and the lower part for the block permutation approach.}
    \label{tab:oA_size}
\end{table}

Larger grid sizes (see Table \ref{tab:oA_40} in the appendix) lead to power values close to $100\%$ in the otherwise same scenarios, even for an anisotropy ratio of $\sqrt{2}$. There are no substantial differences between the two approaches in terms of power then, but block permutation still has the advantage of keeping the significance level even for large ranges. The differences between using the Genton and the Matheron estimators become small as well. Only  the MCD.diff estimator results in lower power values when the anisotropy includes a rotation. 

In summary, for Gaussian data without outliers, the isotropy test which uses the block permutation approach with Matheron's variogram estimator seems to be the best choice. A benefit compared to the subsampling approach is that the test keeps the significance level in all situations investigated here.

\subsection{Contamination with outliers}
\label{ch:sim_out}

Now we investigate the behaviour of the different tests in case of Gaussian data with outliers, considering various outlier distributions and different fractions $\epsilon$ of outliers. Our interest is the Gaussian process $Z(\boldsymbol{s})$ described at the beginning of Section \ref{ch:simulations}, but instead of $Z(\boldsymbol{s})$, we observe the stochastic process $Y(\boldsymbol{s})$ which is contaminated by one of the following two types of outliers. 

The first type are isolated outliers, which occur randomly over the grid. For these, we select the $n_0 = \lceil\epsilon\cdot n \rceil$ outlier positions $N_{n_0}$ uniformly at random from all $n$ locations, with $\epsilon$ being the fraction of outliers. All locations $\boldsymbol{s}$ belonging to $N_{n_0}$ follow the outlier distribution $W(\boldsymbol{s}) \sim\mathcal{N}(\mu_0, \sigma_0^2)$, while all remaining locations follow the process of interest $Z(\boldsymbol{s})$,
\begin{align*}
    Y(\boldsymbol{s})|N_{n_0} = \begin{cases} 
    Z(\boldsymbol{s}),~\text{ if } \boldsymbol{s} \notin N_{n_0} \\
    W(\boldsymbol{s}),~ \text{ if } \boldsymbol{s} \in N_{n_0}.
    \end{cases}
\end{align*}

The second type of outliers occurs spatially aggregated in a block. For each location $\boldsymbol{s}_0 = (x_0, y_0)^T$ we define a block neighbourhood $N(\boldsymbol{s}_0)$ of size $\lceil \epsilon \cdot n \rceil$ with $\boldsymbol{s}_0$ being as central as possible in the block. The block shape is generated randomly and ranges from squares to elongated rectangles. We randomly draw a location $s_0$ from a uniform distribution on all locations $\boldsymbol{s}_1, \ldots, \boldsymbol{s}_n$. This is described by the random variable $U$. The locations in $N(\boldsymbol{s}_0)$ belong to a block of substitutive outliers. The observed process $Y(\boldsymbol{s})$ then is   
\begin{align*}
    Y(\boldsymbol{s})|(U=\boldsymbol{s}_0) = \begin{cases} 
    Z(\boldsymbol{s}),~\text{ if } \boldsymbol{s} \notin N(\boldsymbol{s}_0) \\
    W(\boldsymbol{s}),~ \text{ if } \boldsymbol{s} \in N(\boldsymbol{s}_0).
    \end{cases}
\end{align*}.

\begin{figure}[H]
    \centering
    \includegraphics[width=1\linewidth]{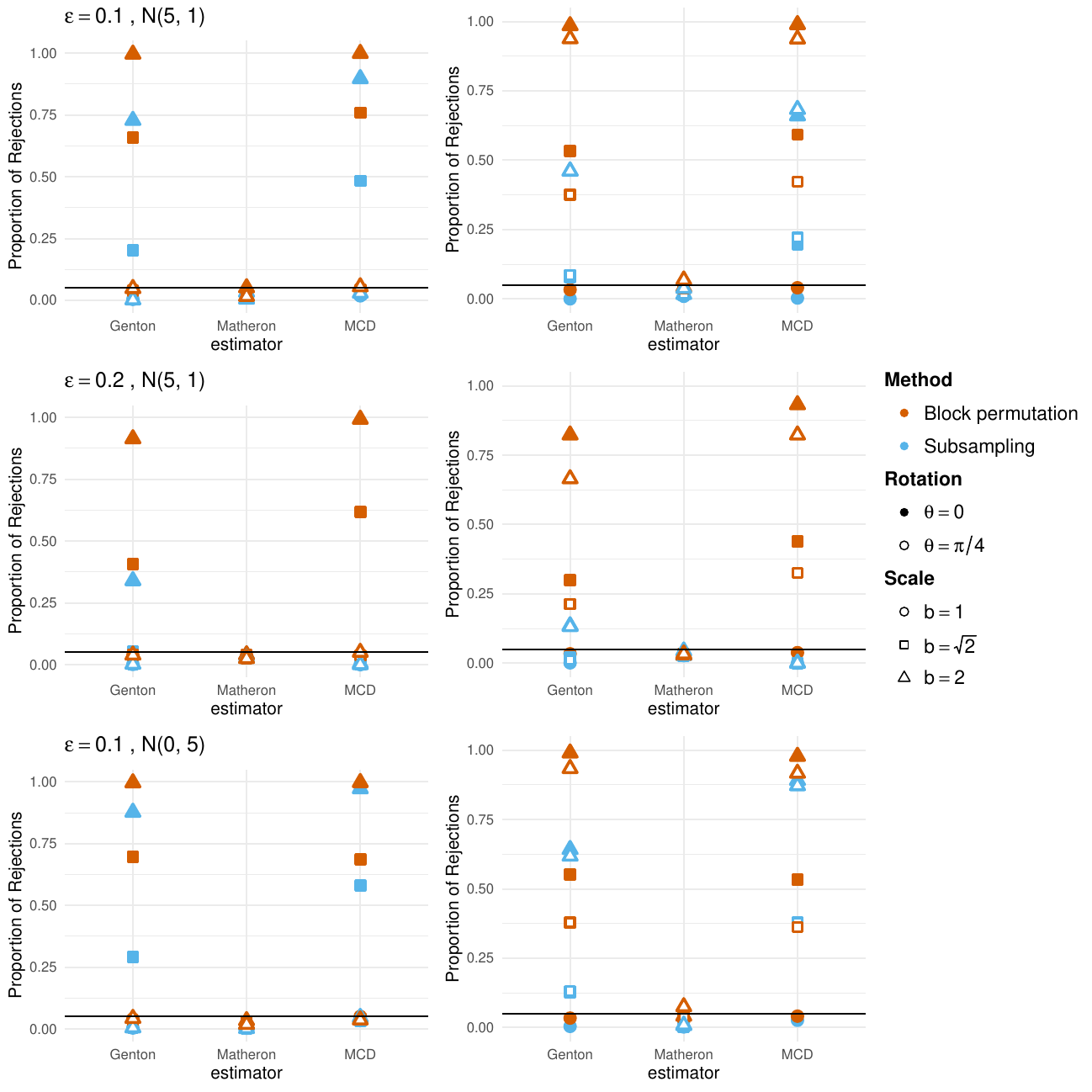}
    \caption{Empirical rejection rates for several fractions of isolated outliers
    with different outlier distributions in
    Gaussian data with the spherical variogram of range 5. In the first column the lag set $\Lambda_1$ is used and in the second column the set $\Lambda_2$. The black line is the significance level $5\%$.}
    \label{fig:IA}
\end{figure}

\begin{figure}[H]
    \centering
    \includegraphics[width=1\linewidth]{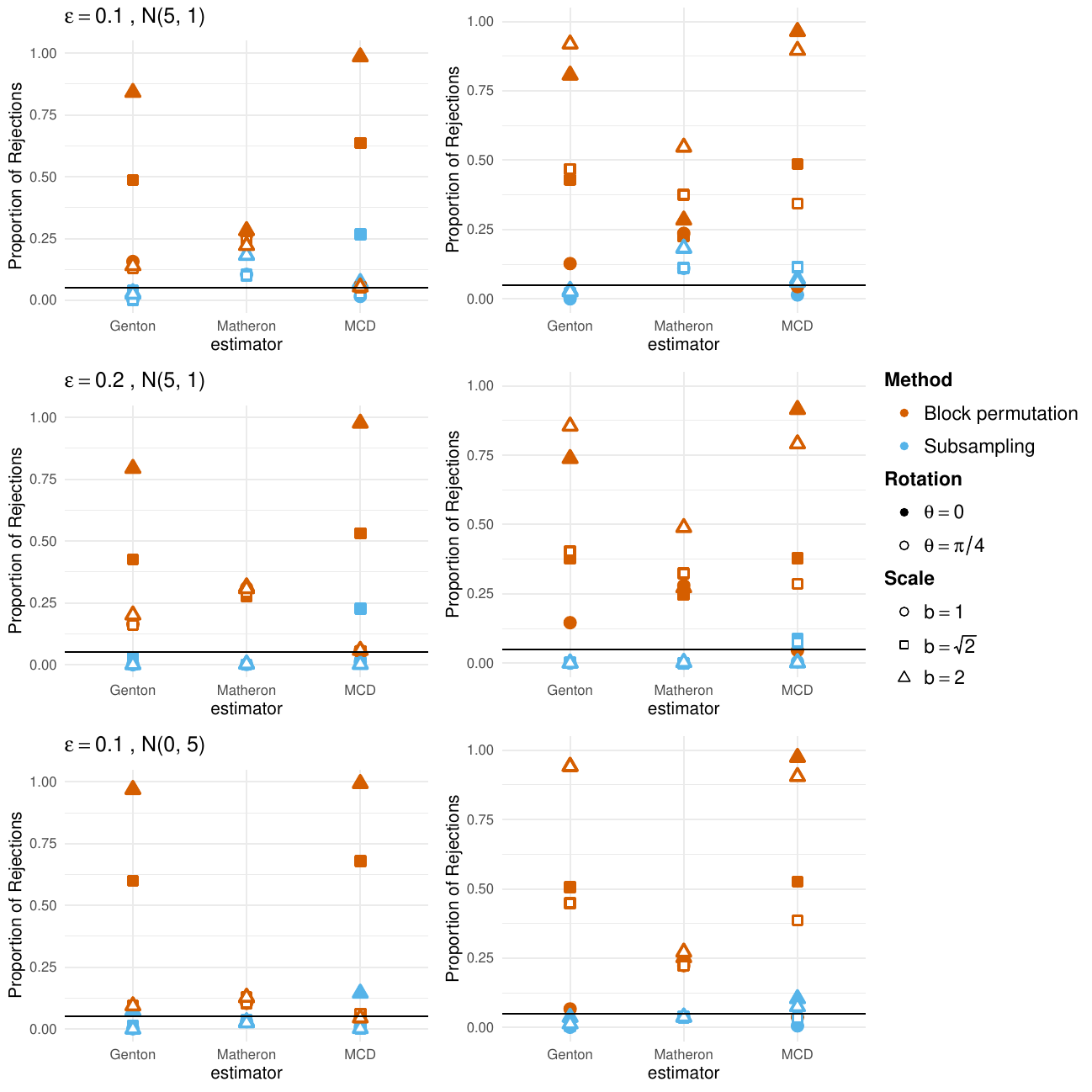}
    \caption{Empirical rejection rates for Gaussian data with block outliers of randomly shapes and the spherical variogram with range 5 in case of several outlier distributions and outlier proportions. In the first column the lag set $\Lambda_1$ is used and in the second column the set $\Lambda_2$. The black line is the significance level $5\%$.}
    \label{fig:BA}
\end{figure}

Figure \ref{fig:IA} shows the simulation results for isolated outliers and Figure \ref{fig:BA} for block outliers. In both figures the spherical variogram with a medium range of $5$ is used and three combinations of outlier distribution and outlier proportion are investigated 
($\epsilon=0.1,~ \mathcal{N}(5,1)$; $\epsilon = 0.2,~ \mathcal{N}(5,1)$; $\epsilon = 0.1,~ \mathcal{N}(0,5)$) 
for the two lag sets
$\Lambda_1$, $\Lambda_2$. We use only these two lag sets here because, in the situations considered here, $\Lambda_3$ performed worse than the other two lag sets even for Gaussian random fields without outliers (see Section \ref{ch:sim_oA}).

The Matheron estimator is known to be non-robust \citep{Genton1998a}. This explains why the tests based on the Matheron estimator are  not robust to either type of outlier. In case of isolated outliers it keeps the level but offers no power. Outlier blocks reduce the power of the resulting tests and increase their type one error.

Furthermore, we observe that the block permutation approach is more robust than the subsampling approach. 
If the outliers occur isolated, the subsampling approach leads to notably less power than the block permutation approach, even when the robust estimators are used. For subsampling, the MCD.diff estimator shows the best performance, but it also loses power if $\epsilon = 0.2$. If outliers occur in a block, the subsampling tests have no power regardless of which estimator is used.

Looking in more detail at the results for isolated outliers and comparing the two robust estimators for the block permutation approach, the differences are mostly small. In contrast, for block outliers, the Genton estimator leads to difficulties in controlling the significance level and results in lower power. The MCD.diff estimator combined with block permutation in such scenarios achieves the highest power while maintaining the nominal significance. As in Section \ref{ch:sim_oA}, the combination  of the MCD.diff estimator and the block permutation approach also fails in the presence of an outlier block for $\Lambda_3$ (see Figures \ref{fig:IA2} and \ref{fig:BA2} in the appendix).

\begin{figure}[H]
    \centering
    \includegraphics[width=1\linewidth]{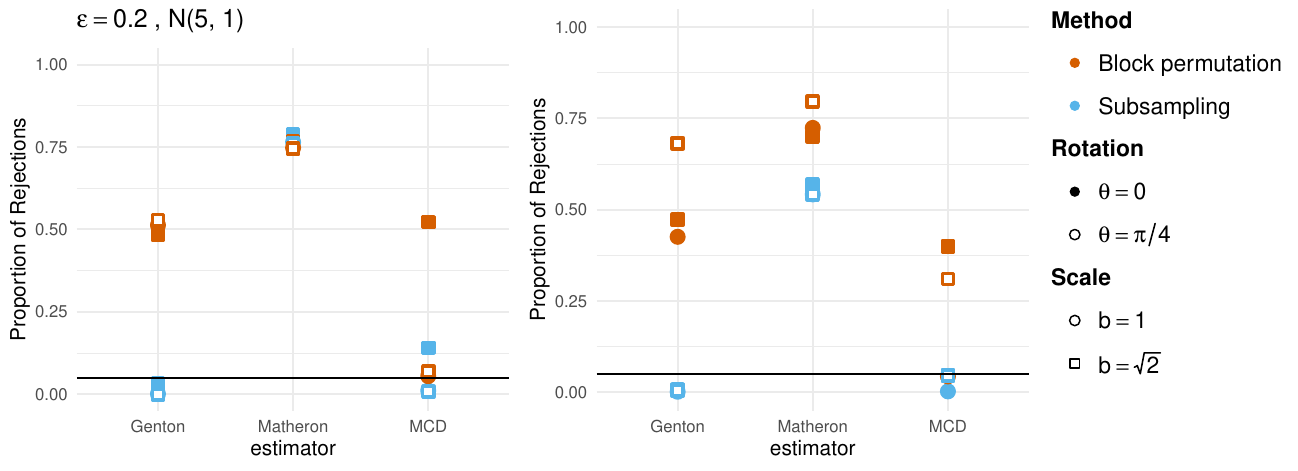}
    \caption{Empirical rejection rates for Gaussian data with elongated rectangular block outliers and the spherical variogram with range 5. In the first column the lag set $\Lambda_1$ is used and in the second column the set $\Lambda_2$. The black line is the significance level $5\%$.}
    \label{fig:BA_rec}
\end{figure}

The superiority of the MCD.diff estimator in combination with block permutation becomes even more evident when focusing on elongated rectangular outlier blocks. The empirical rejection rates for such outliers are shown in Figure \ref{fig:BA_rec}. Again, a spherical variogram with range $5$ is used, a contamination rate of $20\%$ is assumed, and the outliers follow an $\mathcal{N}(0,5)$ distribution. Such an outlier block induces many outliers in one direction and only a few  in the other direction. As a result, the block permutation approach combined with the Genton estimator exhibits some power but does not control the significance level, with type I error rates around $50\%$. In contrast, when the MCD.diff estimator is used with block permutation, the test maintains the nominal level while still achieving reasonable power.
As opposed to this, nearly quadratic outlier blocks lead to results similar to those obtained under random block designs (see Figure \ref{fig:BA_quad} in the Appendix). 

In summary, using the block permutation approach with one of the robust variogram estimators leads to a robust isotropy test. If the suspected shape of the outlier block is an elongated rectangle, the MCD.diff estimator is to be preferred.

\section{Application to Satellite Data}\label{ch:applications}

Landsat 8 is part of the NASA Landsat programme and collects data on Earth’s land surface across nine spectral bands in the visible and shortwave infrared regions \citep{Knight2014}. 
In this application, we use open-access data acquired by Landsat 8 on 20 October 2016, corresponding to a single time point, for a small region of the Brazilian Amazon rainforest.
The region is defined by $x$-coordinates ranging from -7332135 and -7319535 and $y$-coordinates from -1011642 and -995142 in the EPSG:3857 coordinate reference system.  It comprises a grid of 60 × 60 pixels, each representing an area of 30 × 30 metres. All tests for isotropy and variogram estimators considered in the following assume intrinsic or weak stationarity. To make these assumptions plausible, we restrict our attention to a relatively small and homogeneous region consisting solely of a forrested area. The data are publicly available and can be obtained from \url{https://earthexplorer.usgs.gov/}.

Our primary objective here is to assess whether the assumption of isotropy holds for the Normalised Difference Vegetation Index (NDVI). Therefore, we first estimate the variogram in different directions using robust and non-robust variogram estimators, and in a next step we perform the isotropy tests on the data. The NDVI is a widely used vegetation index that quantifies the greenness of biomass. It is calculated from two spectral bands and takes values in the range from -1 to 1, with higher values indicating denser and healthier vegetation \citep{Myneni1995, Tucker1979}.

\begin{figure}[H]
    \centering
    \includegraphics[width=0.5\linewidth]{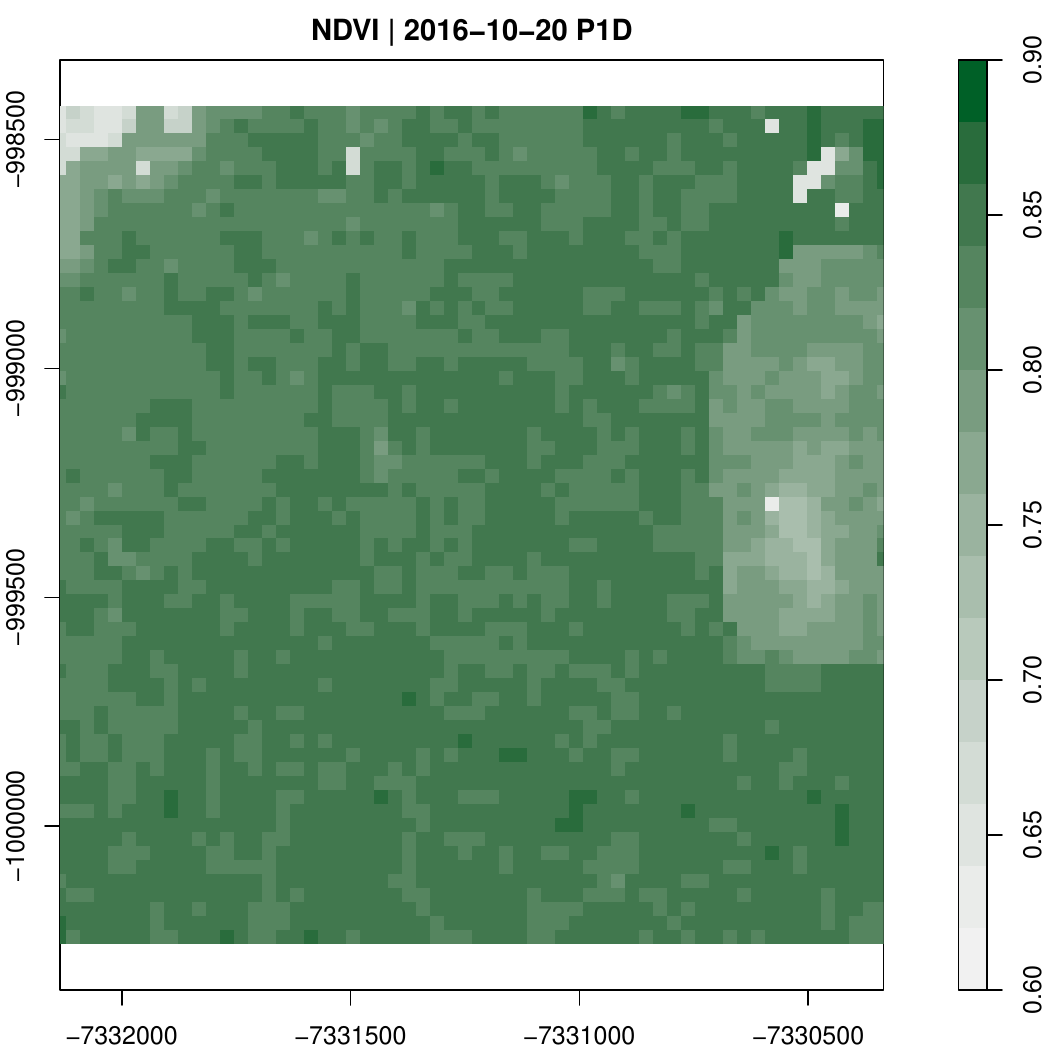}
    \caption{NDVI index in a small region in the Brazilian amazon forest on October 20, 2016 with contamination due to clouds on the right hand side and at the top}
    \label{fig:ndvi}
\end{figure}

The data are illustrated in Figure \ref{fig:ndvi}. Overall, the NDVI values are fairly homogeneous across most pixels, resulting in a relatively low variance. However, two distinct regions, one on the right-hand side and another near the top of the image, exhibit noticeably lower NDVI values compared to the surrounding area. In addition to the spectral bands, Landsat data also include a quality assessment band that provides information on possible contamination of pixels, for example, by clouds, cloud shadows or snow, as well as identifying clear observations \citep{U.S.GeologicalSurvey2019}. For pixels displaying different NDVI behaviour in Figure \ref{fig:ndvi}, the quality band indicates the presence of clouds or cloud shadows, while all remaining pixels are classified as clear. These regions on the right-hand side and at the top can therefore be interpreted as blocks of outliers. Since the variability in the data is very small (with a median absolute deviation of approximately $0.009$), we standardise the data using this scale measure to obtain variogram estimates on a more interpretable scale.

\begin{figure}[H]
    \centering
    \includegraphics[width=0.8\linewidth]{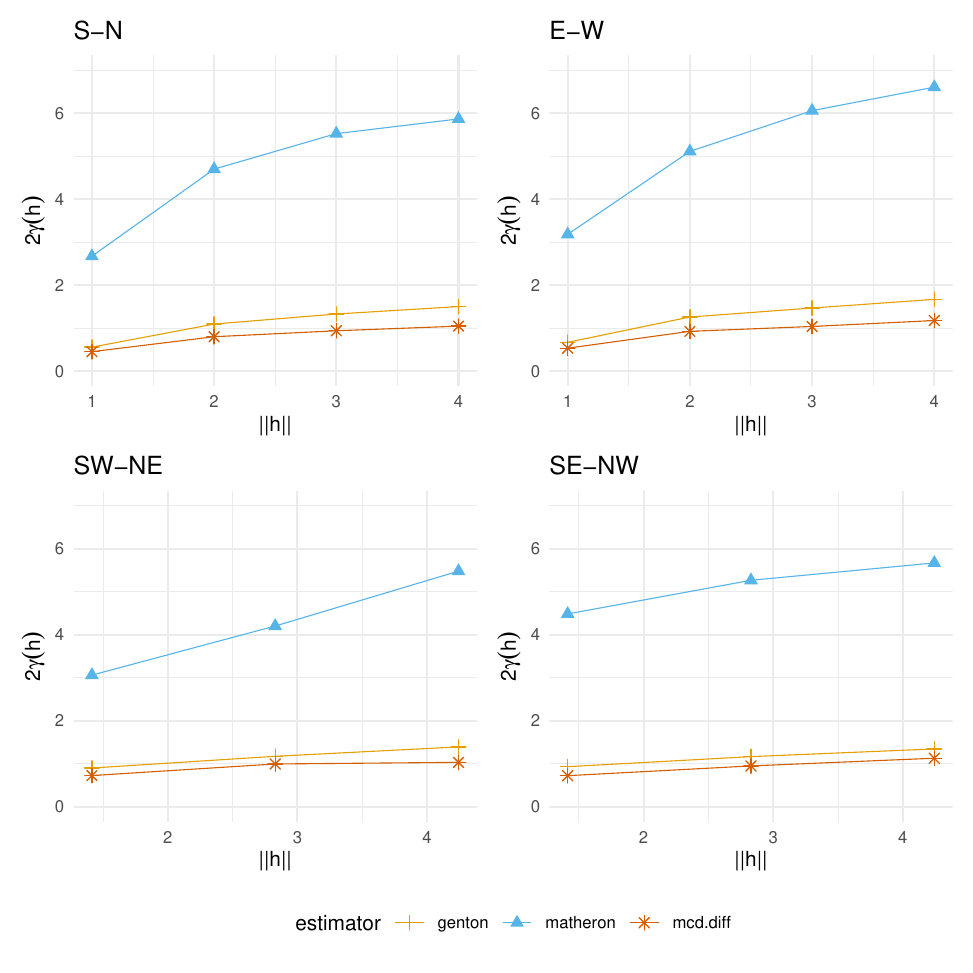}
    \caption{Estimated variogram using different estimators in four different directions for a small region with $60 \times 60$ pixels of the Brazilian amazon forest for data with block outliers.}
    \label{fig:varog-ndvi}
\end{figure}

We estimate directional variograms using the Matheron estimator, the Genton estimator, and the reweighted MCD.diff variogram estimator. For the east–west and south–north directions, we set $h_{\max} = 4$, while for the remaining two directions we use $h_{\max} = 3$. The estimated variograms are displayed in Figure \ref{fig:varog-ndvi}.  
The Matheron variogram estimator is strongly influenced by outliers, which is reflected in clearly higher estimates. Both robust estimators show very similar results. The figure shows no differences between the estimates for the southwest to northeast and the southeast to northwest directions. The differences between the south-north and east-west directions are minor. The estimates for the east-west direction are slightly higher (see also Table \ref{tab:application}). The graphs show no clear evidence of anisotropy. In a next step, the isotropy tests described in this paper are applied at a significance level of $5\%$. The results of the tests can be found in Table \ref{tab:application}. In addition to the p-values of the tests, the table also includes the variogram estimates for the lag vectors used in the test. For the block permutation test all p-values, except the one using MCD.diff and $\Lambda_3$, are small and lead to rejection of the null hypothesis of isotropy at a significance level $5\%$. The p-values of the subsampling tests are all slightly  higher. This means that some variants of the subsampling test reject the null hypothesis, while others do not. 

\begin{table}[ht]
    \centering
        \begin{tabular}{c|ccc|ccc}
        & \multicolumn{6}{c}{test results: p vaules} \\
        & \multicolumn{3}{c}{Subsampling} & \multicolumn{3}{c}{Block permutation} \\
        estimator & $\Lambda_1$ & $\Lambda_2$ & $\Lambda_3$ & $\Lambda_1$ & $\Lambda_2$ & $\Lambda_3$ \\ \hline
        Matheron & 0.06 & 0.03 & 0.05 & 0.02 & 0.00 & 0.00 \\
        Genton & 0.04 & 0.09 & 0.12 & 0.00 & 0.01 & 0.02 \\
        MCD.diff & 0.03 & 0.05 & 0.09 & 0.01 & 0.01 &  0.11
        \end{tabular}
        
        \vspace{0.2cm}
        \begin{tabular}{c|cccccccc}
         & \multicolumn{8}{c}{variogram estimation for lag vector $\boldsymbol{h}$}  \\
        estimator & $(1,0)^T$ & $(0,1)^T$ & $(1,1)^T$ & $(1,-1)^T$ & $(1,2)^T$ & $(-2,1)^T$ & $(2,1)^T$ & $(-1,2)^T$ \\ \hline
        Matheron & 3.18 & 2.67 & 3.06 & 4.48 & 4.21 & 5.67 & 4.60 & 5.23  \\
        Genton & 0.67 & 0.56 & 0.90 & 0.93 & 1.19 & 1.34 & 1.31 & 1.17  \\
        MCD.diff & 0.56 & 0.45 & 0.76 & 0.74 & 0.98 & 1.03 & 1.03 & 0.91 
        \end{tabular}

             \caption{Variogram estimations for different lags and p-values of the subsampling and block permutation isotropy test using three different estimators for the satellite data}
    \label{tab:application}
\end{table}

\begin{figure}[ht]
\begin{minipage}{0.49\textwidth}
     \includegraphics[width=1\linewidth]{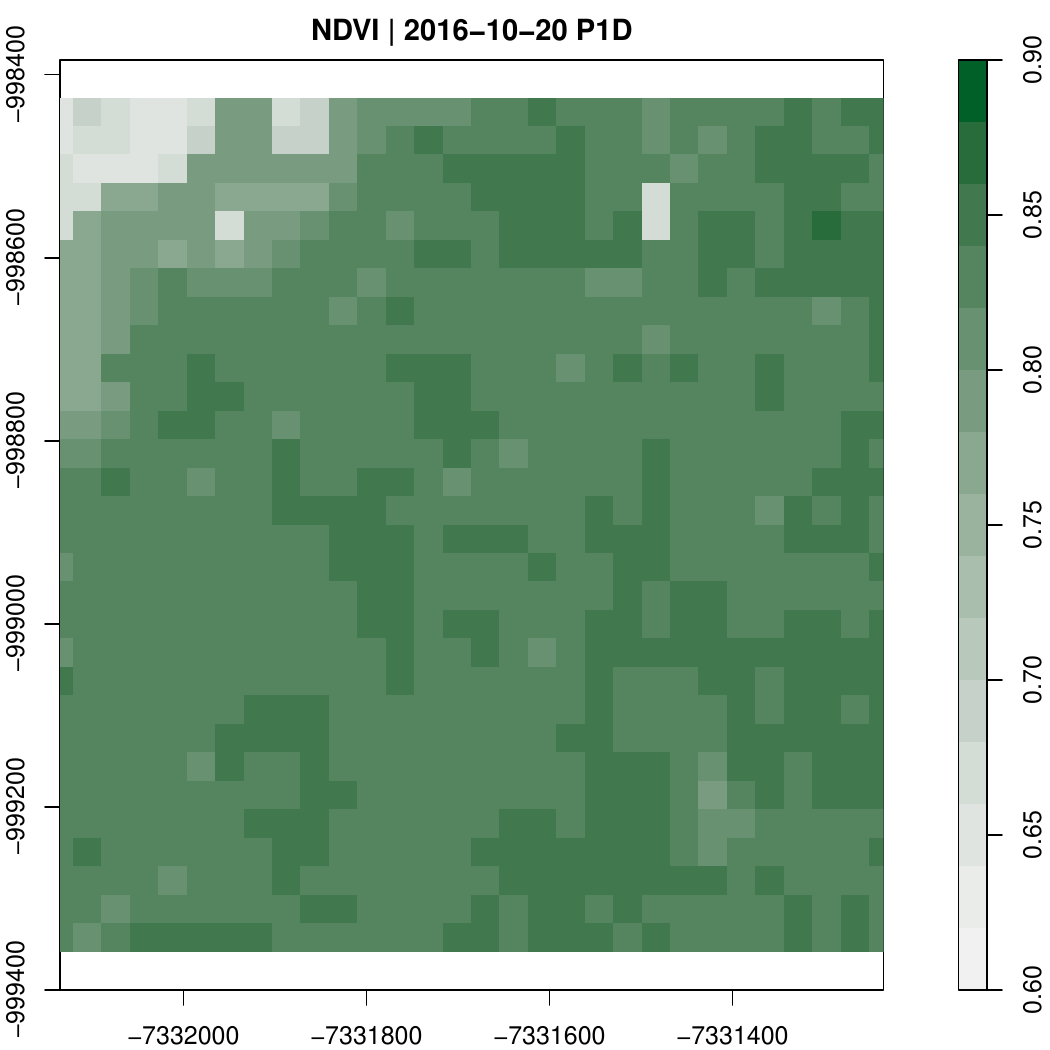}
\end{minipage}
\begin{minipage}{0.49\textwidth}
     \includegraphics[width=1\linewidth]{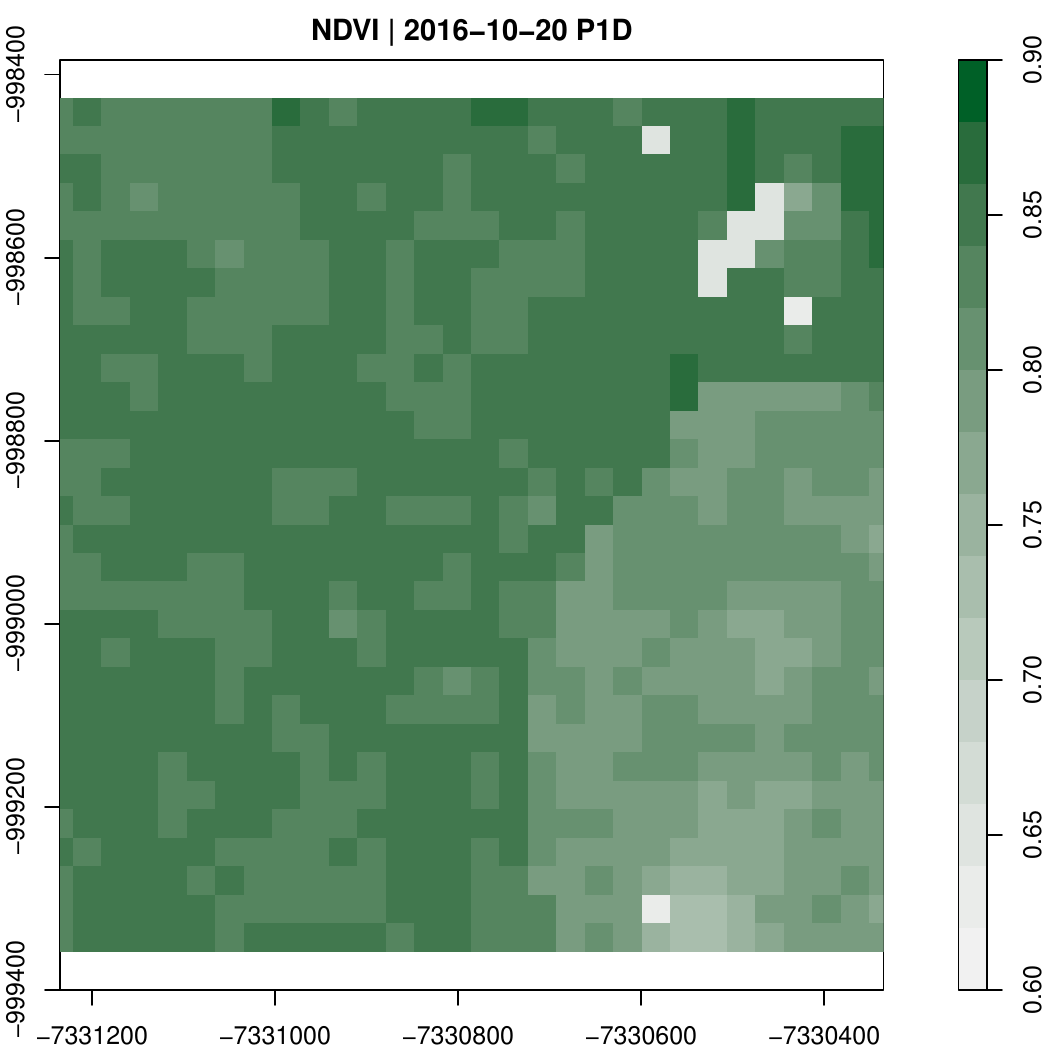}
\end{minipage} \\
\begin{minipage}{0.49\textwidth}
     \includegraphics[width=1\linewidth]{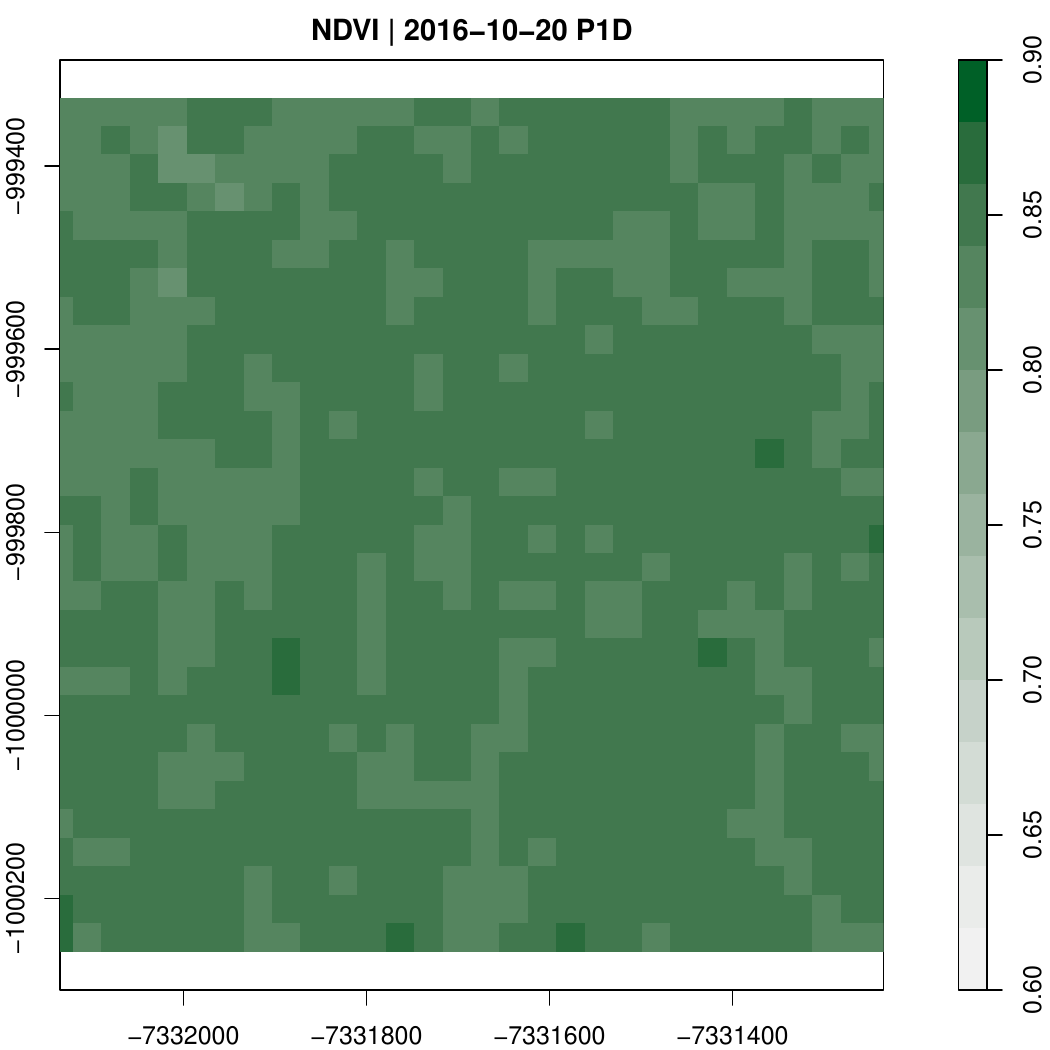}
\end{minipage}
\begin{minipage}{0.49\textwidth}
     \includegraphics[width=1\linewidth]{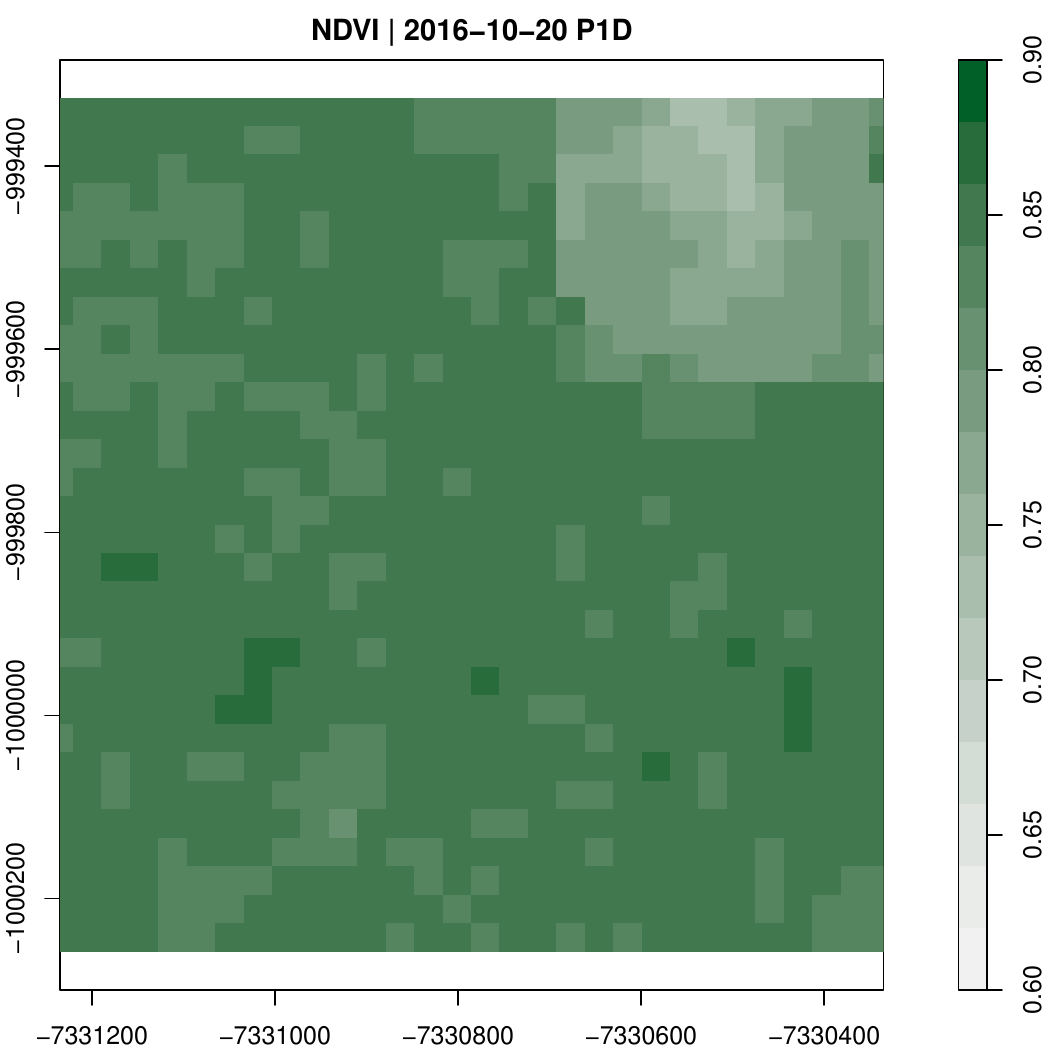}
\end{minipage}
    \caption{NDVI index in four small regions in the Brazilian amazon forest on October 20, 2016. The dataset on the bottom left contains no outliers, while the three other dataset are contaminated with outlier blocks due to clouds.}
    \label{fig:NDVI-4}
\end{figure}

In the next step, the entire $60\times 60$ grid is divided into four $30\times 30$ grids to compare the different tests in various scenarios (with and without outliers, different proportions of outliers, different locations of the outlier blocks). The four data sets are displayed in Figure \ref{fig:NDVI-4}. The grid in the bottom left contains no outliers. The three remaining grids are contaminated with outlier blocks of different sizes due to clouds. The data set on the top left contains approximately $19\%$ outliers, the data set on the bottom right approximately $12\%$, and the data set on the top right approximately $38\%$. The p-values of the isotropy test for the four data sets are given in Table \ref{tab:pvalues4}.

The data set on the bottom left contains no outliers. Comparing the p-values of the different tests for this data set, we observe that, when using the Matheron estimator, the null hypothesis of isotropy cannot be rejected. The same holds for the Genton estimator, which only leads to rejection for $\Lambda_1$ combined with block permutation. In contrast, when using the MCD.diff estimator, the tests reject the null hypothesis in most cases, except when combined with block permutation for $\Lambda_3$. So, there is no clear tendency here. 

The other three data sets contain blocks of outliers. The data set on the bottom right includes a nearly quadratic block of outliers, with approximately $12\%$ of the observations affected. For this data set, none of the tests rejects the null hypothesis of isotropy. For quadratic outlier blocks, all tests maintain the nominal significance level (see Figure \ref{fig:BA_quad} in the Appendix). Hence, this data set does not contradict the assumption of isotropy. 

In contrast, the data set on the top left contains a nearly triangular block of outliers, with about $19\%$ of the data affected. Using robust estimators in combination with block permutation, the tests reject the null hypothesis for $\Lambda_1$ and $\Lambda_2$, while all other tests do not reject it. As discussed in Section 4, the presence of outlier blocks can cause the subsampling test, as well as the tests using the Matheron estimator, to lose power. For the Genton estimator, performance under block outliers depends on the shape of the blocks. Overall, the top-left data set provides some evidence in favour of anisotropy. 

The largest proportion of outliers (approximately $38\%$) occurs in the top-right data set, which contains a large elongated block of outliers, primarily aligned in the south–north direction, with fewer outliers in the east–west direction. Consequently, more differences in the south–north direction are affected by the outliers than in the east–west direction. As a result, variogram estimates in the south–north direction are more heavily influenced by outliers. In Figure \ref{fig:BA_rec} in Section \ref{ch:sim_out}, we observed that this leads to tests based on the Matheron and Genton estimators having power but failing to maintain the nominal significance level. In contrast, block permutation combined with the MCD.diff estimator maintains the significance level and exhibits power (expect for $\Lambda_3$). These tests reject the null hypothesis for $\Lambda_1$ and $\Lambda_2$. Therefore, this data set also supports the presence of anisotropy. 

In summary, the different $30 \times 30$ data sets yield varying test results. This raises doubts whether the assumption of stationarity is justified. Overall, however, there are indications of slight anisotropy between the south–north and east–west directions.

\begin{table}[ht]
      \begin{tabular}{c|ccc|ccc}
        & \multicolumn{6}{c}{p vaules: top left} \\
        & \multicolumn{3}{c}{Subsampling} & \multicolumn{3}{c}{Block permutation} \\
        estimator & $\Lambda_1$ & $\Lambda_2$ & $\Lambda_3$ & $\Lambda_1$ & $\Lambda_2$ & $\Lambda_3$ \\ \hline
        Matheron & 0.08 & 0.08 & 0.12 & 0.27 & 0.00 & 0.14 \\
        Genton & 0.11 & 0.16 & 0.16 & 0.01 & 0.04 & 0.12 \\
        MCD.diff & 0.06 & 0.09 & 0.10 & 0.01 & 0.01 &  0.13
        \end{tabular}

        \vspace{0.2cm}
      \begin{tabular}{c|ccc|ccc}
        & \multicolumn{6}{c}{p vaules: top right} \\
        & \multicolumn{3}{c}{Subsampling} & \multicolumn{3}{c}{Block permutation} \\
        estimator & $\Lambda_1$ & $\Lambda_2$ & $\Lambda_3$ & $\Lambda_1$ & $\Lambda_2$ & $\Lambda_3$ \\ \hline
        Matheron & 0.08 & 0.07 & 0.08 & 0.00 & 0.00 & 0.00 \\
        Genton & 0.08 & 0.11 & 0.13 & 0.02 & 0.05 & 0.17 \\
        MCD.diff & 0.02 & 0.03 & 0.16 & 0.03 & 0.05 &  0.40
        \end{tabular}

         \vspace{0.2cm}
      \begin{tabular}{c|ccc|ccc}
        & \multicolumn{6}{c}{p vaules: bottom left} \\
        & \multicolumn{3}{c}{Subsampling} & \multicolumn{3}{c}{Block
        permutation} \\
        estimator & $\Lambda_1$ & $\Lambda_2$ & $\Lambda_3$ & $\Lambda_1$ & $\Lambda_2$ & $\Lambda_3$ \\ \hline
        Matheron & 0.07 & 0.42 & 0.17 & 0.15 & 0.21 & 0.42 \\
        Genton & 0.06 & 0.15 & 0.13 & 0.03 & 0.07 & 0.21 \\
        MCD.diff & 0.01 & 0.03 & 0.02 & 0.03 & 0.03 &  0.40
        \end{tabular}

         \vspace{0.2cm}
      \begin{tabular}{c|ccc|ccc}
        & \multicolumn{6}{c}{p vaules: bottom right} \\
        & \multicolumn{3}{c}{Subsampling} & \multicolumn{3}{c}{Block
        permutation} \\
        estimator & $\Lambda_1$ & $\Lambda_2$ & $\Lambda_3$ & $\Lambda_1$ & $\Lambda_2$ & $\Lambda_3$ \\ \hline
        Matheron & 0.12 & 0.15 & 0.19 & 0.52 & 0.28 & 0.66 \\
        Genton & 0.52 & 0.78 & 0.53 & 0.49 & 0.75 & 0.88 \\
        MCD.diff & 0.69 & 0.43 & 0.51 & 0.96 & 0.65 &  0.95
        \end{tabular}
    \caption{P-values of the subsampling and block permutation test for isotropy using three different estimators for four different Satellite data sets of size $30\times 30$.}
    \label{tab:pvalues4}
\end{table}

\section{Summary}\label{ch:summary}

A correct specification of the spatial dependence structure of a data set is essential for subsequent analysis steps, such as prediction, as well as for the interpretation of the data. A common simplifying assumption is that of an isotropic spatial dependence structure. However, this assumption should be critically assessed in practice \citep{Guan2004}. 

In this work, we investigate a new robust isotropy test for spatial data on a regular grid. The proposed test is based on the approach of \citet{Guan2004}, which relies on comparing variogram estimates in different directions. We replace the non-robust Matheron variogram estimator \citep{Matheron1962} with two robust alternatives: the Genton estimator \citep{Genton1998a} or the MCD.diff estimator \citep{Gierse2025}. Our construction follows the lines of \citet{Guan2004}, who use subsampling techniques to estimate the covariance of the variogram estimators, as well as to compute p-values for small grids. However, these techniques are not robust, as they rely on estimates from small subgrids. Therefore, we propose a block permutation approach instead, in which the grid is divided into non-overlapping subblocks. A new permuted grid is constructed by randomly rotating some of the blocks by 90 degrees.

We compare the proposed test with that of \citet{Guan2004}. For Gaussian data without outliers, tests based on the Matheron estimator outperform those using robust estimators. Our block permutation approach has the advantage of maintaining the nominal significance level even in the presence of strong spatial dependence. The Matheron estimator is known to be highly sensitive to outliers \citep{Lark2000, Kerry2007, Gierse2025}, which is also reflected in the results for the corresponding tests; therefore, their use cannot be recommended in such settings. The same applies to the subsampling approach, which is also strongly affected by outliers.

In the presence of outliers, we recommend using a robust variogram estimator in combination with the block permutation approach. The Genton estimator performs well if isolated outliers occur, as it yields reliable results across all lag sets. However, if outliers occur in spatially aggregated blocks, the MCD.diff estimator combined with block permutation is preferable. This is particularly true if the size of the outlier block is different for the different directions. In such scenarios, the Genton estimator is not recommended as it leads to inflated Type I error rates.

\bmhead{Acknowledgements}
The authors gratefully acknowledge the computing time provided on the Linux HPC cluster at TU Dortmund University (LiDO3), partially funded in the course of the Large-Scale Equipment Initiative by the German Research Foundation (DFG) as project 271512359. Financial support by the Deutsche Forschungsgemeinschaft (DFG, German Research Foundation; Project-ID 520388526; TRR 391: Spatio-temporal Statistics for the Transition of Energy and Transport) is gratefully acknowledged. 

\bmhead{Supplementary information}
An \texttt{R}-implementation for the simulation study in Section \ref{ch:simulations} as well as the \texttt{R}-Code for Section \ref{ch:applications} is available on GitHub, see \url{https://github.com/JGierse/Robust_Isotropytest.git}.

\bmhead{Data availability statements}
The data are open source data and can be downloaded on \url{https://earthexplorer.usgs.gov/}.  The $x$-coordinates of the region lies between -7332135 and -7319535 and the $y$-coordinates between -1011642 and -995142 in the EPSG:3857 format. 

\bibliography{library}

\begin{appendices}

\newpage

\section*{Appendix}

\begin{figure}[ht]
    \centering
    \includegraphics[width=0.7\linewidth]{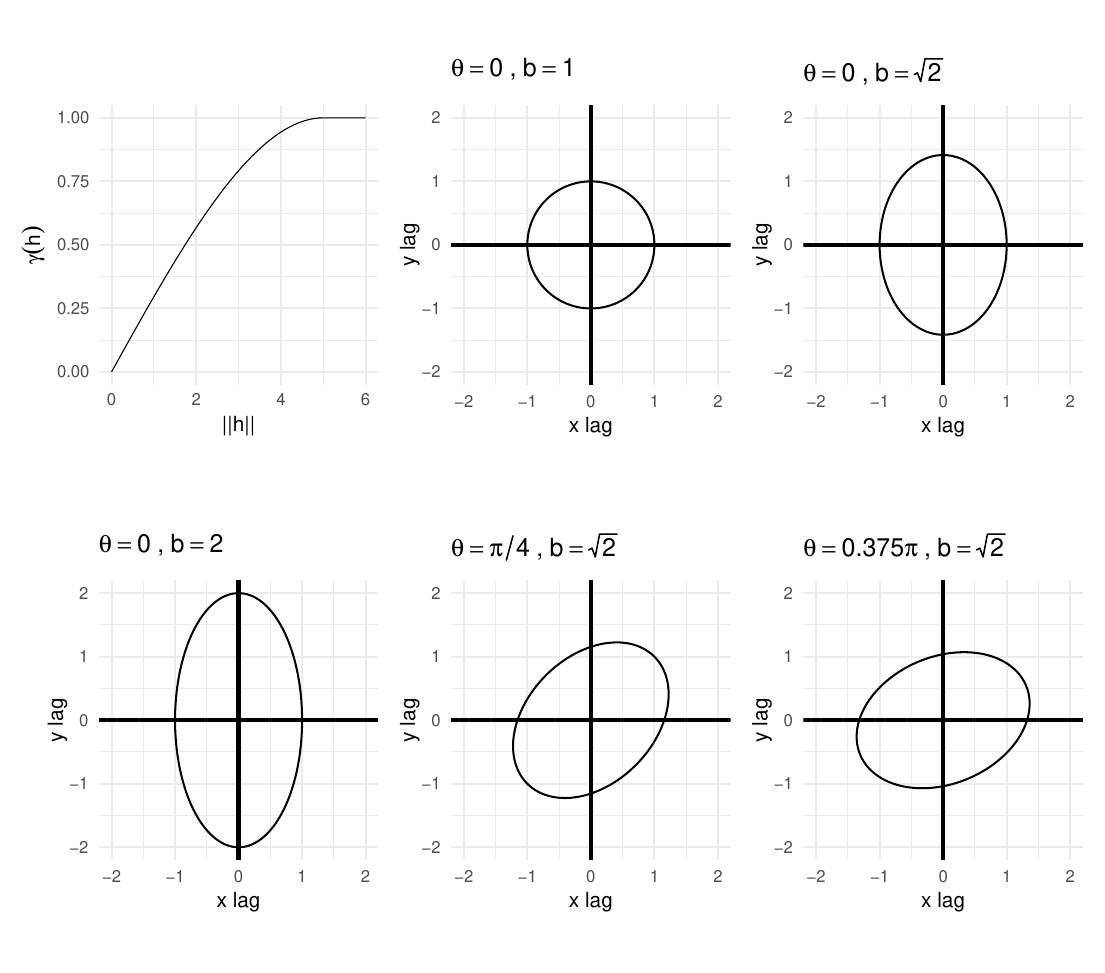}
    \caption{True isotropic variogram with a range of $5$ used in the simulation study, and contours of equal correlation for different angles and ratios of geometric anisotropy. Note that
$\theta = 0$ and $b = 1$ corresponds to an isotropic correlation structure.}
    \label{fig:aniso}
\end{figure}

\begin{table}[ht]
    \centering
    \footnotesize
    \begin{tabular}{cc|rrr:rrr:rrr}
     & &  \multicolumn{9}{c}{Subsampling} \\
      &  & \multicolumn{3}{c}{$\theta = 0; b = 1$} & \multicolumn{3}{c}{$\theta = 0; b = \sqrt{2}$} & \multicolumn{3}{c}{$\theta = \nicefrac{\pi}{4}; b = \sqrt{2}$}  \\
     $r$ & estimator & $\Lambda_1$ & $\Lambda_2$ & $\Lambda_3$ & $\Lambda_1$ & $\Lambda_2$ & $\Lambda_3$ & $\Lambda_1$ & $\Lambda_2$ & $\Lambda_3$  \\ \hline 
     \multirow{3}{*}{2} & Matheron & 4 & 5 & 5 & 100 & 100 & 100 & 5 & 99 & 99  \\
                        & Genton & 2 & 2 & 2 & 100 & 100 & 97 & 3 & 94 & 85 \\
                        & MCD.diff & 9 & 11 & 13 & 99 & 98 & 95 & 9 & 92 & 86  \\ \hdashline
    \multirow{3}{*}{5} & Matheron & 4 & 6 & 4 & 100 & 100 & 99 & 5 & 99 & 95  \\
                       & Genton & 3 & 4 & 4 & 100 & 99 & 96 & 4 & 98 & 89  \\
                       & MCD.diff &  9 & 13 & 15 & 99 & 98 & 96  & 11 & 96 & 93  \\ \hdashline
    \multirow{3}{*}{8} & Matheron & 6 & 7 & 6 & 100 & 100 & 98 & 6 & 99 & 94 \\
                       & Genton & 6 & 10 & 14 & 100 & 99 & 98 & 8 & 99 & 95   \\
                       & MCD.diff & 13 & 24 & 32 & 100 & 99 & 98 & 14 & 98 & 96  \\
    \end{tabular}
    
    \vspace{0.2cm}
          \begin{tabular}{ccc|rrr:rrr:rrr}
     & &  \multicolumn{9}{c}{Block permutation} \\
      &  & &\multicolumn{3}{c}{$\theta = 0; b = 1$} & \multicolumn{3}{c}{$\theta = 0; b = \sqrt{2}$} & \multicolumn{3}{c}{$\theta = \nicefrac{\pi}{4}; b = \sqrt{2}$}  \\
     $r$ & estimator & blocksize & $\Lambda_1$ & $\Lambda_2$ & $\Lambda_3$ & $\Lambda_1$ & $\Lambda_2$ & $\Lambda_3$ & $\Lambda_1$ & $\Lambda_2$ & $\Lambda_3$  \\ \hline 
     \multirow{3}{*}{2} &  \multirow{3}{*}{Matheron}  & $5\times 5$ & 3 & 3 & 1 & 100 & 100 & 100 & 4 & 98 & 99   \\
     & & $8\times 8$ & 4 & 4 & 3 & 100 & 100 & 100  & 5 & 98 & 99 \\
     & & $10\times 10$ & 5 & 5 & 5 & 100 & 100 & 99 & 5 & 98 & 98 \\ \addlinespace
                        &  \multirow{3}{*}{Genton}  & $5\times 5$ & 3 & 3 & 1 & 100 & 100 & 100 & 4 & 94 & 96  \\
     & & $8 \times 8$ & 4 & 4 & 4  & 100 & 100 & 99  & 4 & 95 & 96 \\
     & & $10\times 10$ & 5 & 4 & 6  & 100 & 100 & 98 & 5 & 94 & 93  \\ \addlinespace
                        &  \multirow{3}{*}{MCD.diff}  & $5\times 5$ & 3 & 3 & 0 & 98 & 96 & 83 & 4 & 79 & 52 \\
     & & $8 \times 8$ &  4 & 4 & 1  & 98 & 96 & 76  & 5 & 79 & 50 \\
     & & $10\times 10$ &  5 & 5 & 2 & 98 & 94 & 59 & 5 & 77 & 38  \\ \hdashline
     \multirow{3}{*}{5} &  \multirow{3}{*}{Matheron}  & $5\times 5$ & 3 & 3 & 0 & 100 & 100 & 100 & 5 & 98 & 99  \\
     & & $8 \times 8$ &  4 & 4 & 2 & 100 & 100 & 100 & 6 & 98 & 98  \\
     & & $10\times 10$ & 4 & 4 & 4 & 100 & 100 & 99  & 5 & 96 & 96  \\ \addlinespace
                        &  \multirow{3}{*}{Genton}  & $5\times 5$ & 4 & 3 & 1 & 100 & 100 & 99 & 4 & 97 & 93   \\
     & & $8 \times 8$ & 3 & 4 & 3 & 100 & 100 & 98 & 5 & 96 & 94  \\
     & & $10\times 10$ & 4 & 4 & 5 & 100 & 100 & 97 & 6 & 95 & 90 \\ \addlinespace
                        &  \multirow{3}{*}{MCD.diff}  & $5\times 5$ & 3 & 3 & 0  & 98 & 97 & 76  & 5 & 88 & 42 \\
     & & $8 \times 8$ & 3 & 4 & 0  & 99 & 96 & 51 & 5 & 87 & 26 \\
     & & $10\times 10$ & 4 & 4 & 1 & 98 & 96 & 28  & 5 & 84 & 12 \\ \hdashline
      \multirow{3}{*}{8} &  \multirow{3}{*}{Matheron}  & $5\times 5$ & 4 & 4 & 2 & 100 & 100 & 100 & 5 & 98 & 99 \\
     & & $8 \times 8$ & 5 & 6 & 5 & 100 & 100 & 100 & 6 & 98 & 98  \\
     & & $10\times 10$ &  6 & 6 & 7 & 100 & 100 & 100 & 6 & 96 & 95 \\ \addlinespace
                        &  \multirow{3}{*}{Genton}  & $5\times 5$ & 4 & 4 & 2 & 100 & 100 & 99 & 5 & 96 & 93   \\
     & & $8 \times 8$ & 5 & 5 & 4 & 100 & 99 & 99 & 5 & 96 & 92  \ \\
     & & $10\times 10$ &  5 & 5 & 5 & 100 & 99 & 97 & 5 & 93 & 87 \\ \addlinespace
                        &  \multirow{3}{*}{MCD.diff}  & $5\times 5$ & 4 & 3 & 0 & 98 & 97 & 73 & 4 & 86 & 39 \\
     & & $8 \times 8$ & 4 & 4 & 1  & 99 & 96 & 47 & 5 & 87 & 21 \\
     & & $10\times 10$ & 4 & 4 & 1  & 98 & 95 & 26  & 5 & 82 & 10 \\
    \end{tabular}
    \caption{Empirical rejection rates (in percent) for Gaussian data without outliers in case of variograms with different ranges $r$ and a $40\times 40$ grid. The first table contains the results for the subsampling approach for subsamples of size $6\times 6$ and the second table for the block permutation approach using different block sizes.}
    \label{tab:oA_40}
\end{table}

\begin{figure}[ht]
    \centering
    \includegraphics[width=\linewidth]{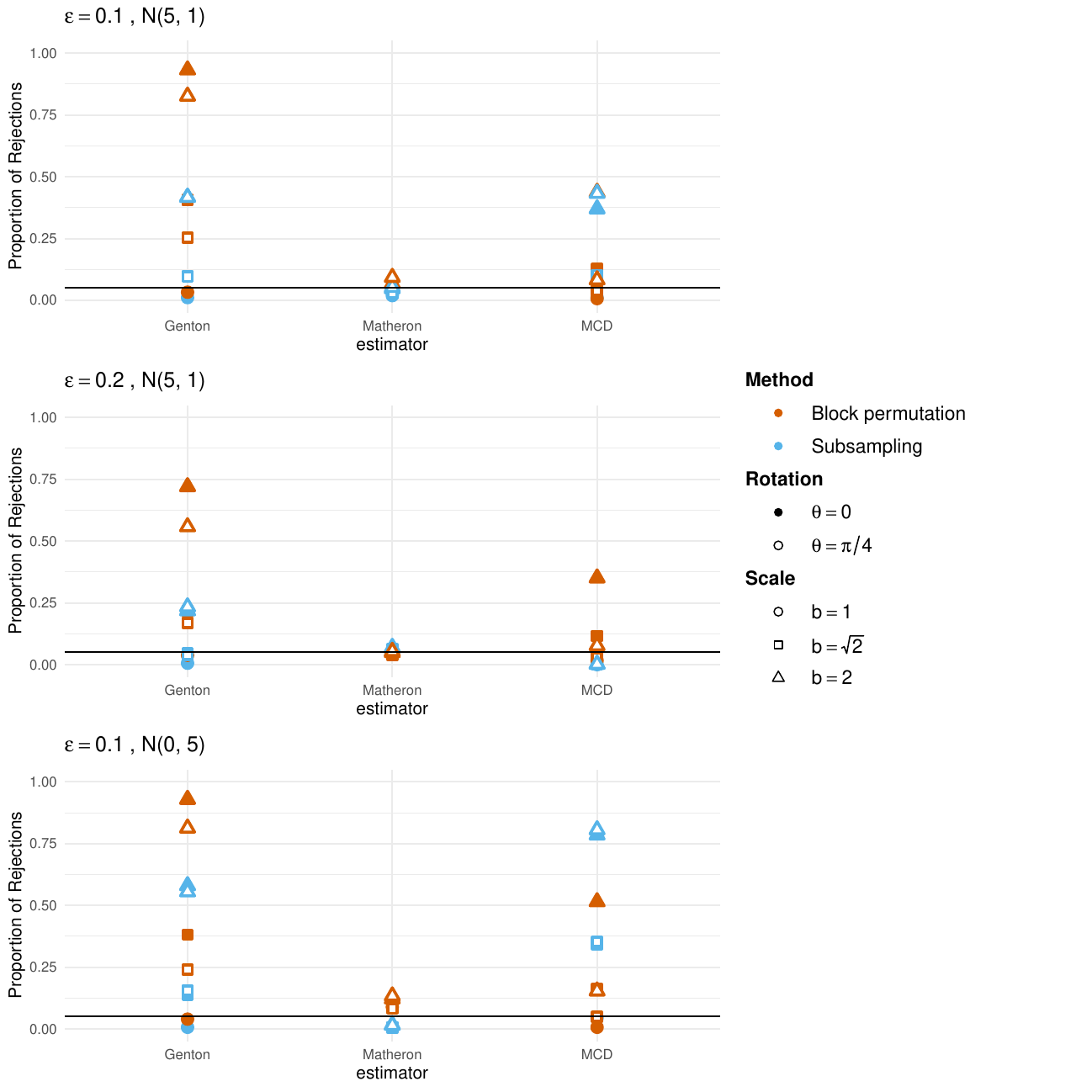}
    \caption{Empirical rejection rates for several fractions of isolated outliers with different outlier distributions in
    Gaussian data with the spherical variogram of range 5. The lag set $\Lambda_3$ is used and the black line is the significance level $5\%$.}
    \label{fig:IA2}
\end{figure}

\begin{figure}[ht]
    \centering
    \includegraphics[width=\linewidth]{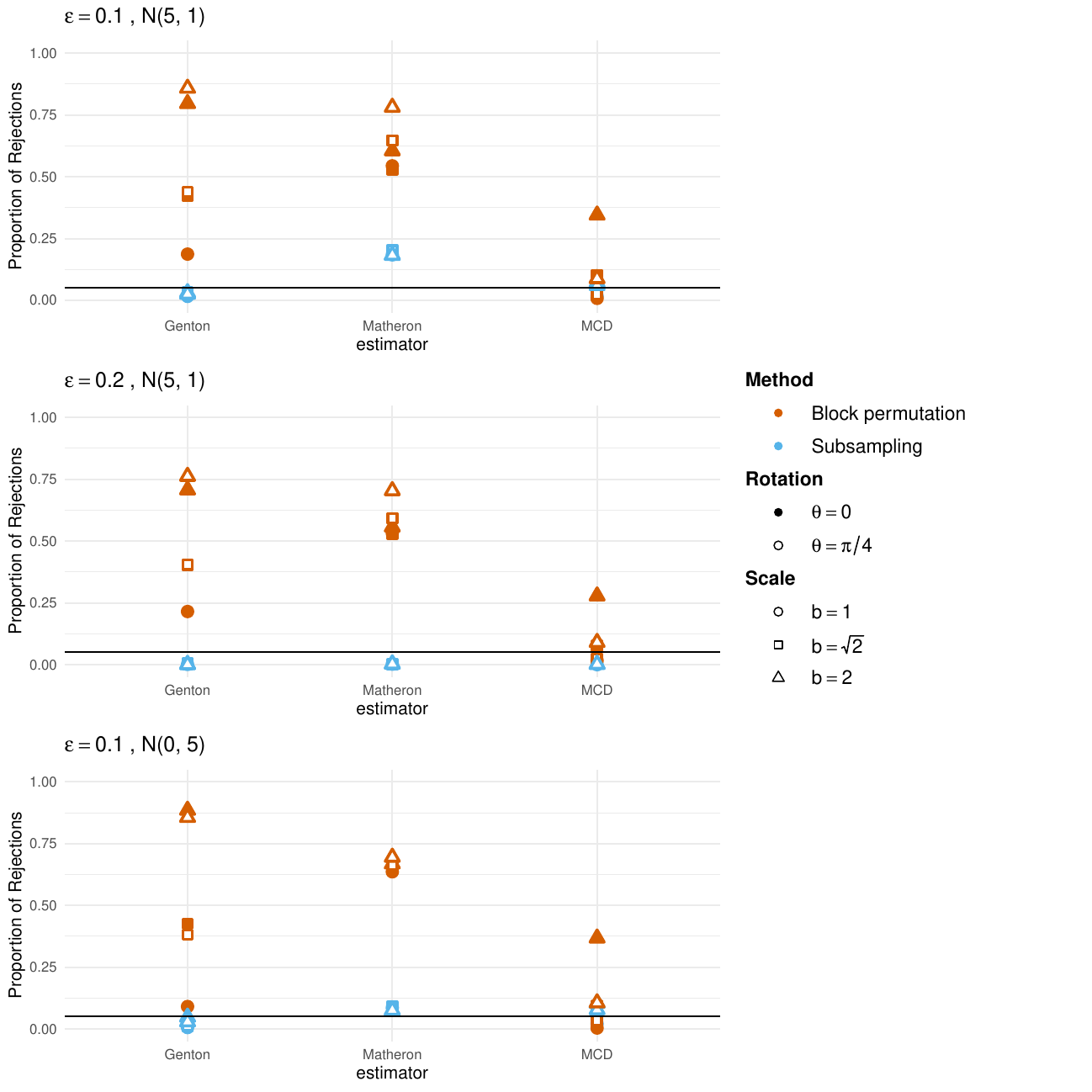}
    \caption{Empirical rejection rates for Gaussian data with block outliers of randomly shapes and the spherical variogram with range 5 in case of several outlier distributions and outlier proportions. The first column the lag set $\Lambda_3$ is used and the black line is the significance level $5\%$.}
    \label{fig:BA2}
\end{figure}

\begin{figure}[ht]
    \centering
    \includegraphics[width=\linewidth]{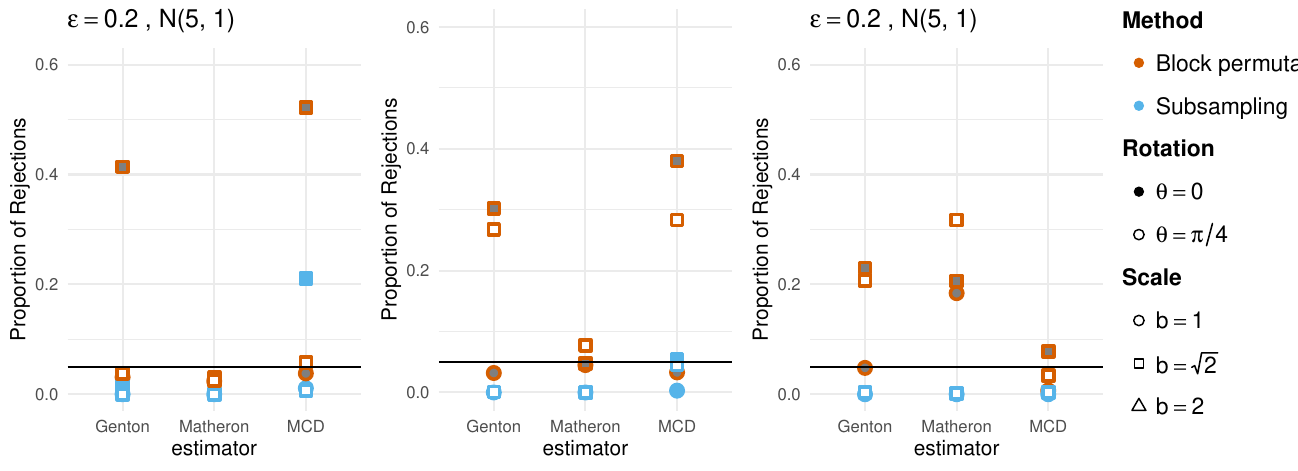}
    \caption{Empirical rejection rates for Gaussian data with rectangular block outliers and a spherical variogram with range 5. In the first column the lag set $\Lambda_1$ is used, in the second column the set $\Lambda_2$ and in the last column the set $\Lambda_3$. The black line is the significance level of $5\%$.}
    \label{fig:BA_quad}
\end{figure}

\end{appendices}

\newpage



\end{document}